\documentclass[preprint]{vgtc}               % preprint
\ifpdf%                                % if we use pdflatex
  \pdfoutput=1\relax                   % create PDFs from pdfLaTeX
  \usepackage{graphicx}                % allow us to embed graphics files
  \DeclareGraphicsExtensions{.pdf,.png,.jpg,.jpeg} % for pdflatex we expect .pdf, .png, or .jpg files
\else%                                 % else we use pure latex
  \usepackage{graphicx}                % allow us to embed graphics files
  \DeclareGraphicsExtensions{.eps}     % for pure latex we expect eps files
\fi%

\graphicspath{{figures/}{pictures/}{images/}{./}} % where to search for the images

\usepackage{microtype}                 % use micro-typography (slightly more compact, better to read)
\PassOptionsToPackage{warn}{textcomp}  % to address font issues with \textrightarrow
\usepackage{textcomp}                  % use better special symbols
\usepackage{mathptmx}                  % use matching math font
\usepackage{times}                     % we use Times as the main font
\usepackage[noadjust]{cite}                      % needed to automatically sort the references
\usepackage{tabu}                      % only used for the table example
\usepackage{booktabs}                  % only used for the table example
\usepackage{amsmath}
\usepackage{tabu}
\usepackage{xcolor, colortbl}
\usepackage{subfigmat}
\usepackage[ruled,linesnumbered]{algorithm2e}
\usepackage{enumitem}
\usepackage{multirow}
\usepackage{scalerel}
\usepackage{soul}
\onlineid{1204}

\vgtccategory{Research}

\vgtcinsertpkg

\title{STULL: Unbiased Online Sampling for Visual Exploration of \\Large Spatiotemporal Data}

\author{
\begin{tabular}{c@{\extracolsep{0.4em}}c@{\extracolsep{0.3em}}c@{\extracolsep{0.4em}}c@{\extracolsep{0.4em}}c} 
Guizhen Wang$^{*}$  & Jingjing Guo$^{*}$ & Mingjie Tang$^{\dag}$ & 
\multicolumn{2}{c}{Jos\'{e} Florencio de Queiroz Neto$^{\ddag}$} \\
Calvin Yau$^{*}$ & Anas Daghistani$^{\S}$ & Morteza Karimzadeh$^{\P}$ & Walid G. Aref$^{*}$ & David S. Ebert$^{\scalebox{0.5}{$\parallel$}}$\\ \rule{0pt}{3ex}   
&\scriptsize Purdue University\thanks{e-mail: \{wang1908, guo49, yauc, aref\}@purdue.edu}
& \scriptsize Chinese Academy of Science\thanks{e-mail: tangrock@gmail.com}
& \scriptsize Federal University of Cear\'{a}\thanks{e-mail: florencio@lia.ufc.br}&\\
& \scriptsize Umm Al-Qura University\thanks{e-mail: ahdaghistani@uqu.edu.sa}
& \scriptsize ~~University of Colorado Boulder\thanks{e-mail: karimzadeh@colorado.edu}&
\scriptsize University of Oklahoma\thanks{e-mail: ebert@ou.edu}&\\
\end{tabular}
}

\teaser{
  \centering
  \includegraphics[width=0.8\linewidth]{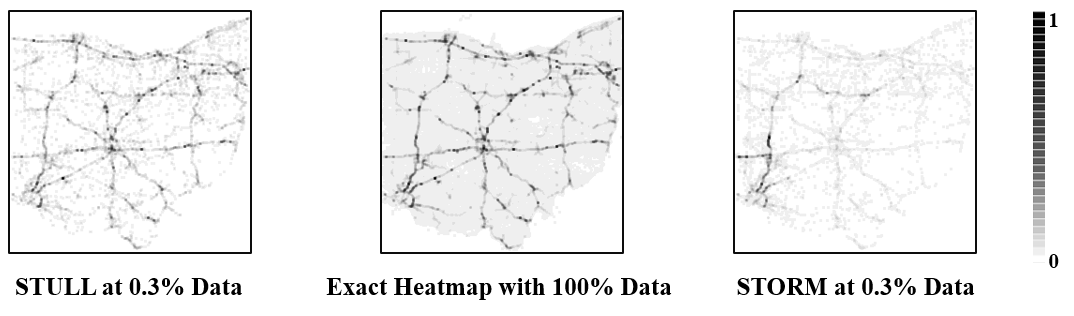}
  \caption{
Visual comparison of Ohio highway traffic incident distributions approximated by 0.3\% data samples retrieved by STULL (left) and STORM (right) in 40 milliseconds or less, against the exact map (middle) at 100\% data, with 32 shades of gray (colorbar).
Both using 0.3\% sample data, the STORM heatmap indicates hotspots are mainly located on the west side of Ohio's highway network, whereas the STULL heatmap shows hotspots across the state and better resembles the exact map.
}
  \label{fig:teaser}
}

\abstract{Online sampling-supported visual analytics is increasingly important,
as it allows users to explore large datasets with acceptable approximate answers at interactive rates.
However, existing online spatiotemporal sampling techniques are often biased, as most researchers have primarily focused on reducing computational latency. 
Biased sampling approaches select data with unequal probabilities and produce results that do not match the exact data distribution, leading end users to incorrect interpretations. 
In this paper, we propose a novel approach to perform unbiased online sampling of large spatiotemporal data.
The proposed approach ensures the same probability of selection to every point that qualifies the specifications of a user's multidimensional query.
To achieve unbiased sampling for accurate representative interactive visualizations, we design a novel data index and an associated sample retrieval plan. 
Our proposed sampling approach is suitable for a wide variety of visual analytics tasks, e.g., tasks that run aggregate queries of spatiotemporal data. 
Extensive experiments confirm the superiority of our approach over a state-of-the-art spatial online sampling technique, demonstrating that within the same computational time,
data samples generated in our approach are at least 50\% more accurate in representing the actual spatial distribution of the data and enable approximate visualizations to present closer visual appearances to the exact ones.
} % end of abstract


\begin{document}
\firstsection{Introduction}

\maketitle

% \section{Introduction}

Online sampling-supported Visual Analytics (VA) allows users to explore large volumes of data at interactive rates when it is not feasible to retrieve and render the whole dataset interactively. 
This is done through continuous retrieval and visualization of retrieved samples that approximate the distribution of the underlying dataset being queried, also known as incremental visualization~\cite{Fisher:trust:2012}.
As users wait for samples to accumulate over time, the sample size increases, which improves the accuracy of the inferred data pattern; this allows users to trade wait time for accuracy~\cite{mozafari2015handbook}.
Therefore, to ensure effective incremental analyses, it is crucial that the progressively retrieved sample is representative of the entire dataset and is not biased. Biased sampling approaches, by definition, sample data with unequal probabilities~\cite{lohr2009sampling}, generate data patterns that deviate from the original dataset, and can lead users to erroneous conclusions (See Figure~\ref{fig:teaser} for an example). 
To ensure trustworthy and reliable data exploration for VA systems, unbiased sampling is critical.

Incremental visualization requires low retrieval latency to support progressive sampling.
In cases using incremental visualization of spatiotemporal data, prevalent spatial sampling approaches~\cite{olken:phdthesis:1993,Chen2020} have slow retrieval times as they use tree-based spatial indexes (e.g., R-Tree~\cite{r-tree:1984}) and iteratively traverse trees from the root to leaf, which often leads to unacceptably high retrieval latency, especially in cases where partial trees reside on disk.
A state-of-the-art spatial online sampling approach, \texttt{STORM}~\cite{STORM_full}, applies sample buffers to tree-based indexes and uses these buffers to substitute high-latency tree traversals.
However, this approach focuses on the efficiency of sample retrieval, without fully resolving the sample bias problem.

In this paper, we present SpatioTemporal Unbiased onLine sampLing (\texttt{STULL}), a novel unbiased online sampling approach that supports incremental visualization and interactive exploration of large spatiotemporal data.
Motivated by the advantages of sample buffers,
\texttt{STULL} proposes a carefully designed sample buffer-based data indexing and sample retrieval plan to ensure that each data point satisfying the user-specified multi-dimensional query has an equal probability of being sampled.
In particular, unlike state-of-the-art spatial online sampling approach~\cite{STORM_full}, our unbiased guarantee is unaffected by the intrinsic spatial distribution pattern of the data.
With our approach, incrementally updated visualizations can not only achieve higher accuracies at the same sample size but also present closer visual appearances to the exact visualizations.
In addition to visual quality,
\texttt{STULL} retrieves samples as efficiently as state-of-the-art approaches (e.g.~\cite{STORM_full}), and allows users to control the number of points sampled through incremental updates.
Through \texttt{STULL}, VA systems can provide unbiased approximate answers to queries for more accurate spatiotemporal visual analytics without adversely impacting the computational performance of the interactive data exploration.
Furthermore, \texttt{STULL} supports sampling both stored data and streaming data, making it suitable for visual analytic environments that leverage both types of data, such as social media analytics tools~\cite{chae2012spatiotemporal, SensePlace3}.

Our experiments confirm the effectiveness and efficiency of \texttt{STULL} in producing unbiased samples for large spatiotemporal data queries.
Compared to the state-of-the-art online spatial sampling approach~\cite{STORM_full} on historical data, in the same computational time, \texttt{STULL} improves the approximate spatial accuracy by at least 50\% when sampling less than 5\% of the original dataset.
Using our approach, approximate visualizations reduce visual differences from visualizations encoding the exact answers.
For streaming data, \texttt{STULL} takes less than 500ms on average to index incoming streaming data (1000$\sim$4000+ tweets per second~\cite{Mercury:2014}) for answering queries, well below the response time thresholds for interactive visualization~\cite{liu2014effects}. 
% m: if your approach is important for streaming data, you need to mention this in your abstract. working on streaming data is a very important contribution. 
%reply: I have no idea about state-of-the-art streaming data.

Our contributions include the following:
\begin{itemize}
    \item a novel, unbiased online sampling approach for VA systems to incrementally present approximate yet reliable interactive analyses,
    % incremental visualization of large spatiotemporal data,
    
     \item theoretical guarantees on the unbiased property of our presented approach.
    
    %  \item experiments verifying and validating the unbiased sampling claims in numerical accuracies and visual effects, and
     
    %  \item extensive discussions of the trade-offs made by \texttt{STULL}.

    % \item Formal computational complexity analysis of \texttt{STULL} and extensive experiments that consistently show the indexing and retrieval of existing and streaming data at interactive rates.
\end{itemize}

% The rest of this paper is structured as follows.
%  We elaborate on the drawbacks of biased sampling for visual analytics in Section~\ref{sec:motivation},
%  review related research in Section~\ref{sec:related_work},
%  detail \texttt{STULL} in Section~\ref{sec:sampling-strategy}, 
%  and provide unbiased guarantee in Section~\ref{sec:stus-unbiased-argument}.
%  In Section~\ref{sec:evaluation}, we explain our experiments and discuss obtained results in Section~\ref{sec:discussion}. 
%  Finally, we conclude the paper and present future directions in Section~\ref{sec:conclusion}. 
\section{Visual Analytics And Sampling Bias} \label{sec:motivation}

\textbf{Unbiased sampling} requires that \textit{each record satisfying the query specification has the same probability of being selected}~\cite{lohr2009sampling}.
Conversely, a sampling approach is considered biased if the probability of each individual record being selected is not equal.

Sampling bias can distort patterns of data and render data exploration ineffective and inaccurate ~\cite{Fisher:2016:BigDataExploration}.
For instance, a common aggregation task in crime analysis is to identify spatial hotspots where the most incidents occur.
Data samples retrieved by unbiased approaches should approximate the hotspot patterns regardless of sample sizes; 
biased approaches may be skewed towards locations outside of the true hotspots and may create false hotspots.

Furthermore, such erroneous interpretations can accumulate throughout the sense-making process.
VA systems often support the interactive exploration of data,  following the information seeking mantra~\cite{Shneiderman:1996}: ``Overview first, zoom and filter, then details on demand.'' 
This practice is common in geospatial analysis, where users often start the exploration by examining data patterns across the overall geographic extent, and then identify locations of interest for further investigation. 
However, if sampling is biased toward specific geographic regions, the visual display at the overview level could already be misleading. As a result, it would then exacerbate the biased selection of relevant regions for further exploration.
% m: as well as biasing the insights retrieved from the exploration process.

Sampling bias also impairs incremental visualization.
Incremental VA systems progressively improve upon approximate answers through three main stages~\cite{Angelini2018}: early, mature and definitive.
% Answers presented in the early stage can indicate exact answers and help users preview whether their analytic activities are on track. 
Answers presented in the early stage can reflect the exact answers, helping users evaluate whether or not their analytic activities are on track.
The results of the mature stage can approximate exact answers with acceptable errors and are useful for time-critical tasks. Finally, the definitive stage approximates answers that do not change significantly and can address analytic tasks that require smaller error margins.
However, with the same sample size, answers constructed from biased samples are often further from the exact answers than their unbiased counterparts.
Thus, biased sampling hinders the advent of each stage and prolongs wait time for users.
Moreover, users' trust in approximate answers is an intrinsic challenge of incremental visualization~\cite{Fisher:trust:users:2017} and sampling bias can exacerbate the trust issue.
For example, one effective visualization technique to help users become confident in the analytic results is to compute the exact answers offline so users can compare their selected, approximate answers against exact ones and redo their analyses if needed~\cite{Fisher:trust:2017}. Biased approximate answers can increase the number of times a user has to redo analyses, decreasing the rate at which they can complete tasks.

% \textcolor{orange}{add a real-world story?}

Therefore, for data exploration in VA systems, it is vital for sampling to be unbiased.
This ensures that 
% the sampled data accurately represent the entire result set and 
approximate visualizations can reliably represent the exact answers.
\section{Related Work} \label{sec:related_work}
We organize the state-of-the-art work related to \textsf{STULL} by the following four topics:

\textbf{Interactive exploration of large data:} 
In a VA system, an additional 500-ms computational latency significantly decreased users' enthusiasm for exploring data~\cite{liu2014effects}.
Here we categorize popular techniques that enable VA systems to rapidly process data.
First, well-designed data indexes can avoid selecting most query-irrelevant data, which significantly reduces latency~\cite{Magdy:2014ig,Tanahashi:2015kl}.
Second, data-cube oriented approaches aggregate the original dataset into a hierarchical knowledge graph, and retrieving answers from such a compact graph is efficient for aggregate queries (e.g., imMens~\cite{imMens}, Nanocube~\cite{Nanocubes}, Hashedcubes~\cite{Hashedcubes}, TOPKUBE~\cite{topkube2018}, SmartCube~\cite{SmartCube2020}).
Third, computational latencies can be reduced by
computational parallelism~\cite{SpatialHadoop:2015, VisReduce} or hidden by pre-fetching~\cite{prefetch_hanrahan, prefetch_sigmod16}.
% reduces latencies by simultaneously processing multiple data chunks but falls short if the data volume grows exponentially.
% Next, pre-fetching~\cite{prefetch_hanrahan, prefetch_sigmod16} predicted queries and conducted the corresponding computation beforehand to hide the latency.
Finally, unlike the above techniques, which process the whole dataset to  produce exact results but are slow to return results if the data volume grows exponentially, sampling-based approximate query processing (AQP) techniques~\cite{BlinkDB:2013,DAQ:2015,Databaselearning:2017} use less data to approximate the original dataset but are able to process large volumes of data without performance degradation. This allows VA systems to quickly process small samples of data and produce error-bounded answers, regardless of the total data volume ~\cite{Fisher:2016:BigDataExploration,Mozafari:2017:SIGMOD}. 

\textbf{Sampling-based AQP and sampling bias:}
% BlinkDB~\cite{BlinkDB:2013} ran stratified sampling in parallel to reduce latency.
% In environments where AQP engines process enormous amount of queries, knowledge from previous data queries was saved to improve performance of future sampling~\cite{DAQ:2015,Databaselearning:2017}.
A broad range of applications (e.g., Geosciences~\cite{sampling:geoscience:2003}, Ecology~\cite{sampling:ecology:2014} and population census~\cite{lohr2009sampling}) employ sub-samples of large data to extrapolate characteristics of the whole dataset.
These extrapolations often assume that their input data can equally represent a large geographical distribution, as biases systematically favor partial data and cause overestimation or underestimation of certain data~\cite{sampling:bias:Schadt:2013}.
Research shows that in logistic regression-based classification problems, inconsistency between the whole dataset and sub-sampled data regarding the multi-class data distribution reduced the predictability of trained models~\cite{Oommen2011SamplingBA}.
In ecological research, species distribution models are prone to overfitting, as sample data is biased in favor of regions where it is easy to collect data~\cite{BORIA201473,sampling:bias:Schadt:2013}. 
Thus, unbiased sampling is essential in many domains to better reflect the true state of the world.
To date, unbiased sampling approaches have been widely investigated~\cite{survey_spatial_sampling_2012, lohr2009sampling}.
For example, unbiased graph sampling~\cite{Wu2017,Leskovec2006} considers graph properties (e.g., degree of nodes in social networks) and designs special sampling strategies to ensure the properties extracted from the samples are representative of the whole. 
Our work focuses on unbiased sampling of spatial data in the online manner, which requires the low latency of sampling large volumes of data.

Visualization can affect sampling strategies as well.
For example, some spatial visualization tasks might include more data from low-density areas so that visualizations do not appear to ignore these regions~\cite{visualization-aware-sampling:2016}.
As for scatterplots, in order to keep desired information (e.g., a data outlier) sampled or to avoid overdraw in high-density areas, some approaches~\cite{Chen2014,Wei2020,Hu2020,Chen2020} assign each point an uneven probability of being selected for rendering.
This also occurs in some spatial data analyses~\cite{DasSarma:2012} where each point has intrinsic priority in the sampling procedure (e.g., advertisement ranking).
In essence, the aforementioned sampling approaches use intentional bias to preserve desired visual properties, whereas our approach focuses on visualization scenarios that retain the distribution of the underlying data and support interactive exploration of large data.

\textbf{Online sampling and incremental visual analytics:}
Online sampling approaches~\cite{OLAP, Hellerstein:1997:OA} select data in a continuous way so that VA systems can produce immediate outputs and incrementally refine them. 
Here we review three common types of methods supporting users to conduct progressive analyses.
% , state-of-the-art approaches improve in different aspects.
First, users' demands for analytical accuracy were typically expressed as certain statistical measurements to configure the number of needed samples (e.g., Confidence Interval~\cite{Hellerstein:1997:OA, Fisher:trust:2012, Fisher:trust:users:2017, Sample+Seek}).
Second, randomness is crucial to data sampling, which inevitably causes analytical results generated in previous executions to have some degree of numerical difference from those in subsequent executions.
Consequently, users have difficulty choosing trustworthy answers.
A series of visual analytics approaches~\cite{Fisher:trust:2017,Kim:2015,INCVISAGE:2017} were developed to help users reduce uncertainty and determine the best answers.  
% Pangloss created a visual analytics environment that enabled users to validate their choices with accurate answers afterwards~\cite{Fisher:trust:2017}.
% To reduce perception workload caused by randomness,
% \texttt{IFOCUS} controlled the number of samples for VA systems when updating related results so that crucial visual properties in views (e.g., bar charts) remain consistent with the ordering of the actual values~\cite{Kim:2015}.
% Likewise, I{\small NC}V{\small ISAGE}~\cite{INCVISAGE:2017} smoothly and consistently refined visualizations designed for 1D or 2D {\sffamily Group-by} data queries.
Finally, in addition to mathematical measurements, online sampling processes can also factor in users' perception and measure approximate answer accuracy in terms of perceived information~\cite{PFunk-H:2016, Perception-aware:dsia:2015}.
% For instance, since humans cannot discern minor differences between colors, the latest research works measured perceived information after each incremental visualization refinement so that users had no need to sample more data if it remained visually discernible.~\cite{PFunk-H:2016, Perception-aware:dsia:2015}. 
In this paper, we focus on sampling bias, which hinders visual analytic activities.
Our approach samples spatiotemporal data in an unbiased way and will therefore reduce the inaccuracy of approximate answers.

\textbf{Online sampling of spatial data:} 
Particular sampling techniques build on the special characteristics of spatial data.
Olken et al.~\cite{olken:phdthesis:1993, Olken1995} presented a suite of spatial sampling methods (e.g., \texttt{RandomPath}~\cite{STORM_full, Olken1995}) that conduct back-and-forth traversals over typical hierarchical spatial structures (e.g., R-tree~\cite{r-tree:1984}, Quad-tree~\cite{Quad-tree:1974}) to retrieve samples.
Likewise, similar sampling strategies have been used for object movement trajectories~\cite{Li:2015, TrajectoryTrafficSamping:2001, trajstore:2010:icde} and scatterplots~\cite{Chen2020}.
These approaches traverse their index from root to leaf to obtain one or a few points.
The sampling procedure repeats to retrieve more samples.
% However, the aforementioned approaches must traverse their tree indexes to select one or more points. 
As a result, the sampling time of these methods scales as the number of tree traversals increases. This problem is compounded in cases where the available memory cannot store trees completely and must save partial trees to disks.
% In the worst circumstance where all the tree leaves reside on disk, \texttt{RandomPath} needs at least one disk I/O for a single point.
Disk I/O is much more expensive than in-memory access~\cite{Fujitsu:2011}. Consequently, retrieving samples from disks cannot satisfy time-critical performance requirements that are mandatory in online scenarios~\cite{liu2014effects}.
However, in the big data era, it is common to use hard drives as secondary storage to alleviate memory shortage~\cite{STORM_full}.
To adapt to the hybrid data storage system that allocates data to both memory and hard drives, \texttt{STORM}~\cite{STORM_full} proposed a novel data index that uses sample buffers to substitute expensive tree traversals. 
Sample buffers pack well-selected sample points into disk blocks~\cite{BufferTree:2003}.
Batched disk I/O operations can quickly load a significant number of disk blocks into memory.
As such, loading these buffers from disks is more efficient than traditional tree-based approaches.
To the best of our knowledge, \texttt{STORM} is the first approach to employ online sampling of spatial data~\cite{STORM_full}.
% Unlike the approach focusing on sampling efficiency, 
Section~\ref{sec:intuition} elaborates details regarding the two sampling genres.
In this paper, we focus on sampling bias arising from the sampling procedure equipped with sample buffers.
Sampling bias can be avoided either by ensuring equal sampling probability for each point or by involving remedies to correct the bias.~\cite{survey_spatial_sampling_2012}. 
Our approach avoids bias by ensuring each point has the same probability of being selected, and can conduct unbiased sampling of discrete spatiotemporal data records, while satisfying the latency requirement for interactivity.

\section{STULL}
\label{sec:sampling-strategy}
 
As users issue queries, \texttt{STULL} continuously samples data so that the VA system can create rapid visualizations and progressively improve them. 
% (Figure~\ref{fig:stull_and_va}).
During a  single incremental update, \texttt{STULL} retrieves sample points per the spatial and temporal specifications of a particular query.
After receiving samples, the visualization side generates and updates the visuals.
This section introduces the computational details of \texttt{STULL}.
Section~\ref{sec:intuition} gives some intuition on the advantage of our sampling strategy.
Sections~\ref{sec:stus-index-structure},~\ref{sec:stus-index-construction} and~\ref{sec:stus-append-data} detail the scalable data index design, creation and updating. 
Section~\ref{sec:stus-data-query} provides details of the sample retrieval procedure with guaranteed unbiasedness. 

% \begin{figure}[!htb] 
%   \centering
%   \includegraphics[width=1\linewidth]{figures/STULL_VA.png}
%   \caption{Incremental visualization workflow leveraged by STULL.}
%   \label{fig:stull_and_va}
% \end{figure}

\subsection{Intuition}
\label{sec:intuition}

Efficient sample retrieval is crucial for online sampling. 
A sampling plan that randomly selects a subset of points from a collection of data points is apparently not efficient because the retrieval accesses all of the points even if users query merely a small part of the data.
A scalable plan involves indexing data and retrieving samples from the index, as the index can minimize the accessed data to the subsets specified by queries.
Specific to spatial sampling, spatial indexes ( e.g., R-tree~\cite{r-tree:1984} and Quad-tree~\cite{Quad-tree:1974}) are widely used to organize data in a spatial hierarchy.
These trees often store all the points into leaf cells.
The sample retrieval procedure (e.g., \texttt{RandomPath}~\cite{Olken1995, STORM_full}) starts from the tree root, randomly chooses a child in terms of some metrics (e.g., the point volume belonging to each child), and recursively picks a child of the chosen cell until it reaches a leaf from which a point is selected.
The same procedure repeats until the desired number of points is collected.
This type of approach produces samples that represent the queried data in an unbiased manner, but the retrieval process is not efficient.
First, the sample retrieval latency increases in proportion to the tree traversal cost.
Second, the cost of traversing trees and selecting data from leaves can exceed the latency bound specified by interactive data exploration~\cite{liu2014effects} if the available memory cannot store the whole index and partial of indexes reside on a disk. 
In cases where data indexes are stored on hard drives, loading non-leaf cells and reached leaves into memory is a lengthy process because each access requires at least one disk Input/Ouput (I/O) operation and completing all the I/Os is time-consuming and at least one to two orders of magnitude slower than in-memory access~\cite{Fujitsu:2011}.
To reduce the number of I/Os, an advanced approach, \texttt{STORM}~\cite{STORM_full}, proposes storing samples satisfying a query specification in a continuous region on the disk. A batch I/O operation can sequentially scan the disk region to retrieve samples, which is faster.
In terms of tree-based spatial hierarchy, \texttt{STORM} allocates a continuous region (known as a buffer~\cite{BufferTree:2003}) on a disk for each of its non-leaf cells.
Each buffer stores data samples that are retrieved in advance from the spatial range represented by its linked cells.
Thus, the points stored in a sub-tree root's sample buffer can approximate all the data indexed by the sub-tree and act as a sample set.
Likewise, the union of points in the sample buffers belonging to the root's children can approximate the data distribution.
As such,  progressively merging more relevant buffers can form an online sampling manner. 
In summary, the recent buffer-based approach can retrieve samples rapidly and support VA systems at interactive rates.

\begin{figure}[!htb] 
  \centering
  \includegraphics[width=1\columnwidth]{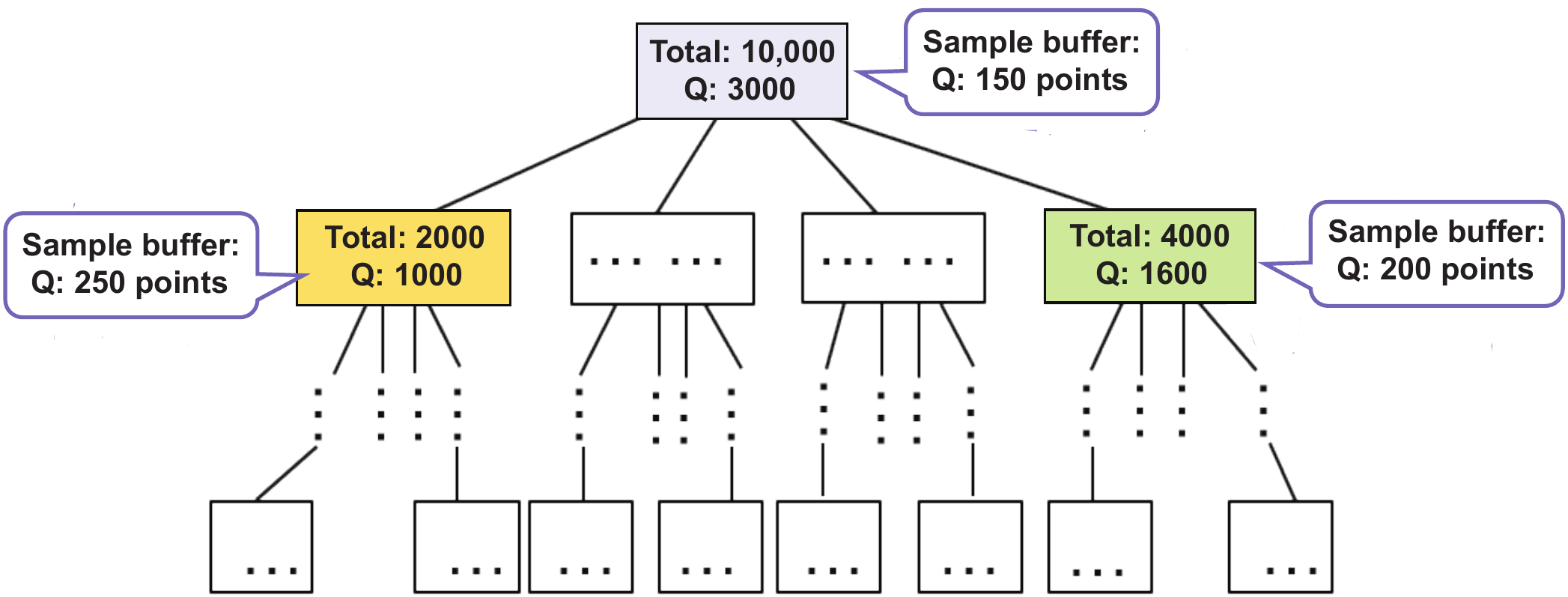}
  \caption{Sampling bias issue in the fixed-sized sample buffer design.
  $Q$ is a query.
  Each sample buffer has 500 random data points.
   Numbers inside each buffer lists the number of points satisfying $Q$. Numbers inside each cell list the total number of points in the spatial range of the cell and the number of points satisfying $Q$ respectively.
  }
  \label{fig:fixed_size_issue}
\end{figure}

Since buffer-based sampling approaches retrieve samples in the units of buffers, the state-of-the-art \textbf{fixed-sized} design proposed by \texttt{STORM}~\cite{STORM_full} in which each buffer has the same number of points raises bias (Figure~\ref{fig:fixed_size_issue}).
In this example, each non-leaf cell has a 500-point sample buffer. At level 2, collectively, the union dataset of the orange and green cells has $43.3\%$ points (i.e. $\frac{1000+1600}{2000+4000}$) satisfying Q, whereas the sample set that is a combination of their sample buffers has 45\% points (i.e., $\frac{250+200}{500+500}$) satisfying Q.
Therefore, the samples cannot accurately approximate the dataset. 

To avoid this issue, we propose a \textbf{proportionally-sized} design.
In this design, each cell's buffer caches $100\alpha$ percent of its data.
The sample buffer is therefore \textit{proportional} to the cell's specified range. 
In the case when a query relates to multiple cells (Figure~\ref{fig:fixed_size_issue}),
the union of these cells' buffers will contain exactly 100$\alpha$ percent of the data being queried.
Therefore, the {proportionally-sized} design can approximate the distribution of the queried data without bias whereas the {fixed-sized}~\cite{STORM_full} is contingent on the spatial index itself.
\texttt{STULL} uses the proportionally-sized design so that it can prevent such issues and ensure that the samples can represent the exact spatial distribution unbiasedly.

% On the contrary, proportional sized buffer design allows each cell to cache $100\alpha$ percent of its data that then aggregates to a sample containing $\alpha$ percent sample from both $Q1$ and $Q2$, leaving the distribution unaltered. 

% A sample buffer of \texttt{STULL} contains $100\alpha$
% %Mingjie: what is the 100\alpha meaning ? 
% percent of the data within its specified spatio-temporal range. 
% The sample buffer is therefore \textit{proportional} to the cell's specified range. 

% Figure \ref{fig:storm} illustrates the issue of the fixed-sized design. 
% In this example, each non-leaf cell has a 500-point sample buffer. At level 2, collectively, the union dataset of the orange and green cells has the distribution of $43.3\%$ (i.e. $\frac{1000+1600}{6000}$) Q1, and 56.7\% Q2, whereas the sample combining two sample buffers has the distribution of 45\% Q1, and 55\% Q2. On the contrary, proportional sized buffer design allows each cell to cache $100\alpha$ percent of its data that then aggregates to a sample containing $\alpha$ percent sample from both $Q1$ and $Q2$, leaving the distribution unaltered. 
% By sampling data proportional to the categories, \texttt{STULL} can prevent such issues and ensure that the original spatial distribution is preserved.

% \subsection{Performance goals}
% \label{sec:performance_goal}

% Incremental visualization allows a VA system to respond rapidly to user queries and provide trustworthy approximate answers in order for users to make decisions~\cite{Fisher:trust:2012, Fisher:trust:2017}.
% Accordingly, centering on response efficiency and answer effectiveness, we extrapolate the following observations used to regulate \texttt{STULL}.

% \begin{enumerate}[label = {\textbf{C\arabic*}}]
%     \item Samples represent queried subsets of the data in an unbiased manner. Therefore, answers derived from these samples cannot over or underestimate the exact answers.
    
%     \item Sample retrieval in one incremental update is on average quick (no more than 500ms~\cite{liu2014effects}) so that visualizations can progressively refine at an interactive rate.
    
%     \item The first response to queries cannot be slow because seeing answers quickly can help users detect any incorrect query specifications and repair them as early as possible~\cite{Fisher:trust:2012}.
%     %m: this point is duplicated with previous one
    
%     \item Specific to spatiotemporal data, retrieval time is scalable to query specifying wider temporal or spatial ranges.
    
%     \item The size of samples retrieved in one update is configurable so that users can  adjust update rates on demand.
% \end{enumerate}

\subsection{Index Design}
\label{sec:stus-index-structure}

\begin{figure}[!htb] 
  \centering
  \includegraphics[width=0.9\linewidth]{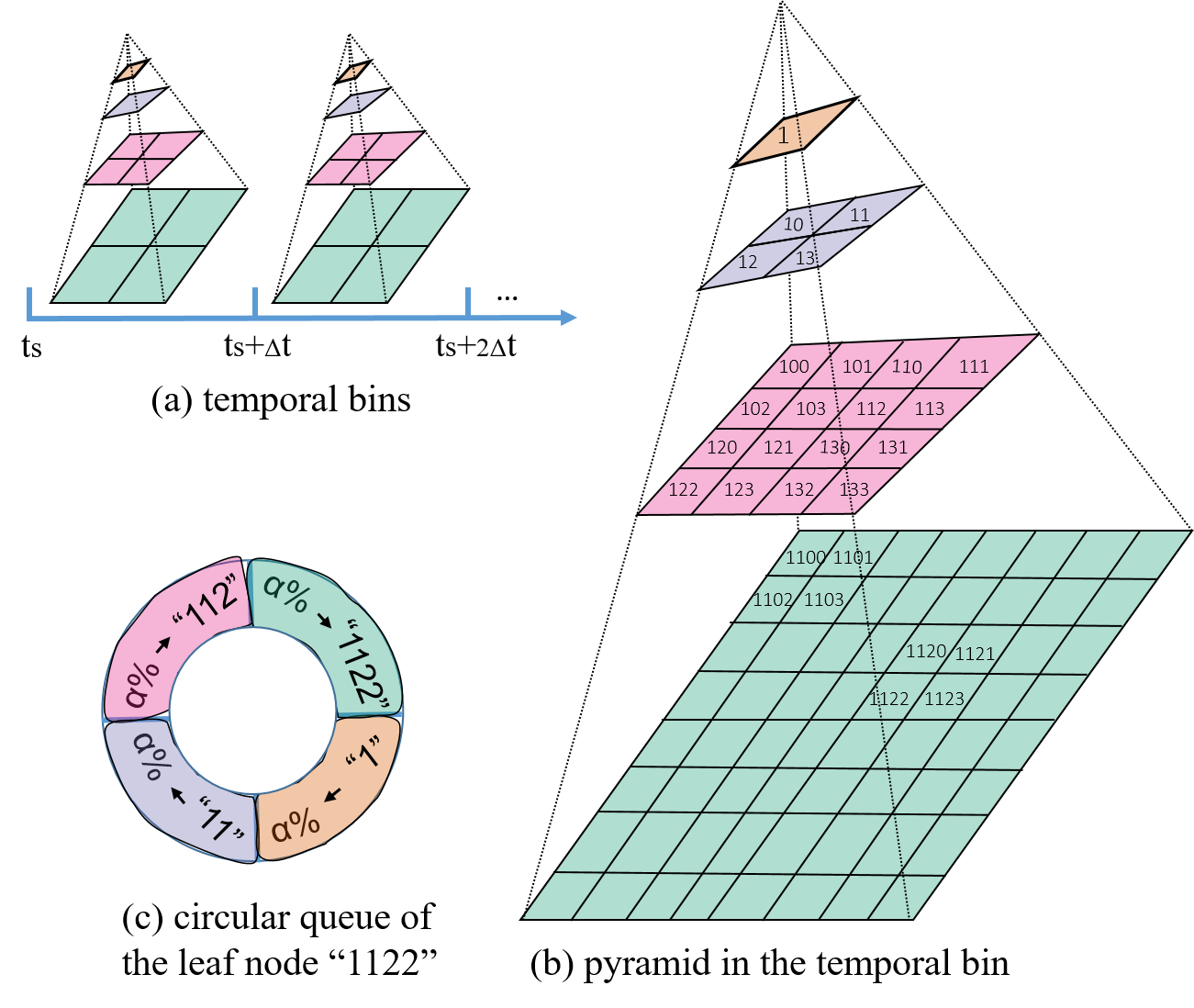}
  \caption{The data index. 
  (a) shows the temporal index beginning at $t_s$.
  Each segment in (a) is a temporal bin. 
  Each temporal bin sets $\alpha=0.25$ and uses a four-level pyramid (in (b)) to spatially organize data.
  A pyramid leaf uses a four-segment circular array (in (c)) to store data.
  Each non-leaf cell has a sample buffer to store data.
  } 
  \label{fig:top-down-data-organization}
\end{figure}

\texttt{STULL} indexes data with an ordered list of pyramids that represents the spatio-temporal segmentation of the data. (Figure~\ref{fig:top-down-data-organization}).  
The temporal range of the data, $\Delta t$, is first divided into adjacent, non-overlapping, equal-sized temporal bins, each indexing a subset of the data that falls into its range.
Within each temporal bin, data is further indexed with a pyramid (e.g., \texttt{Mars}~\cite{Magdy:2014ig}) per its spatial dimensions.
Each pyramid recursively and equally divides the data's spatial range into four fixed-sized rectangular sub-ranges until the $\lceil \frac{1}{\alpha} \rceil$-th level.
$\alpha$ is the reciprocal of a pyramid's height.
Unlike a Quad-Tree, each pyramid's  non-leaf cells have sample buffers. 
These sample buffers follow the \textit{proportionally-sized} design to 
cache $100\alpha$ percent of points randomly selected from their spatiotemporal ranges.
Therefore, each pyramid level has in total $100\alpha$ percent of points.
Accordingly, a pyramid has $\frac{100}{100\alpha} = \frac{1}{\alpha}$ levels in total.

At the bottom level of a pyramid, leaf cells store all of the data within their range in a {circular array} (Figure~\ref{fig:top-down-data-organization}(c)). We divide each circular array into $\frac{1}{\alpha}$ segments in terms of the pyramid height, where each segment contains $100\alpha$ percent of the data. 
These segments will be used to add points into non-leaf cells' sample buffers (Section~\ref{sec:stus-index-construction}) and participate in sample retrieval (Section~\ref{sec:stus-data-query}).
Figure~\ref{fig:top-down-data-organization}(c) exemplifies the circular array in the leaf with the id ``1122'' in Figure~\ref{fig:top-down-data-organization}(b). 
Since $\alpha=0.25$, its circular queue has four segments.

Cells from the non-bottom levels of a pyramid cache randomly selected samples from their respective ranges in one-dimensional arrays, termed \textbf{sample buffers}.
Collectively, data in the sample buffers form a sample that approximates the distribution of the original data (Section~\ref{sec:stus-data-query}).

% \subsection{Data Indexing}
\subsection{Index Creation}
\label{sec:stus-index-construction}

To build the index, \texttt{STULL} first puts each data point in the appropriate temporal bin. 
Then, starting from the root level of the pyramid, in a top-down fashion, we proceed to the appropriate spatially-ranged leaf cell and insert this point into its circular array, and repeat this process for all data points. 
At completion, the circular arrays at the bottom levels of all pyramids will contain the entire data set.

Next, in each pyramid, \texttt{STULL} adds points to sample buffers at the non-bottom levels of pyramids.
First, each leaf cell randomly shuffles data in its circular array (Figure~\ref{fig:top-down-data-organization}(c)).
Second, a bottom-up procedure copies segments of data from leaf cell circular arrays into sample buffers of their ancestor cells.
Take one leaf for example, 
% since a pyramid has $\frac{1}{\alpha}$ levels, 
its circular array has $\frac{1}{\alpha}$ segments. 
%each containing $100\alpha$ percent of the data. %within the individual leaf.
In clockwise order, the data in the first segment is copied into the sample buffer of the root level, data in the second ancestor cell on the second level, and so forth, until the $\frac{1}{\alpha}$-th segment is copied.
% The $\frac{1}{\alpha}$-th segment is on the bottom level.
Algorithm~\ref{algo:build-sample-buffer} shows the pseudo-code of this procedure.

\begin{algorithm} [!htb]
\SetKwData{Left}{left}\SetKwData{This}{this}\SetKwData{Up}{up}
\SetKwFunction{Union}{Union}\SetKwFunction{FindCompress}{FindCompress}
\SetKwInOut{Input}{input}\SetKwInOut{Output}{output}
\Input{A list $T$ consisting of temporal bins that need to build sample buffers}
\BlankLine
\For{each time bin $t$ in $T$}{
empty all sample buffers\;
\For{each leaf $u$ of $t$}{
random shuffle data in $u$'s circular array\;
$ n \leftarrow $ the length of $u$'s circular array\;
$index \leftarrow 0$\;
\For{$i = 1: \frac{1}{\alpha}$}{
$c \leftarrow $ the ancestor cell in the $i$-th level and belonging to the path from $u$ to the root\;
$b \leftarrow $ the whole data in the $i$-th segment of $u$'s circular queue\;
Add $b$ to $c$'s sample buffer\;
% $index \leftarrow index + \alpha n$
}
}
random shuffle all sample buffers\;
}
\caption{Building sample buffers}
\label{algo:build-sample-buffer}
\end{algorithm}\DecMargin{1em}              
%m: the caption would be changed to "build index?"

\subsection{Sample Retrieval}
\label{sec:stus-data-query}

\texttt{STULL} retrieves sample points that satisfy a given query $Q$ in an incremental way.
Suppose a VA system plans to use $100\theta$ percent of data points to generate quick answers and progressively refine answers with the same number of new points.

In order to keep the sampling result unbiased in the temporal dimension, \texttt{STULL} retrieves only $100\theta$ percent of sample points from each temporal bin requested per $Q$, where $\theta$ denotes the ratio of points desired by users.
The union of samples retrieved from all of the requested bins is the set of samples the visualization side uses for visual computation.

Retrieving $100\theta$ percent of points from a single temporal bin takes the following steps.
First, \texttt{STULL} randomly picks a pyramid level $l_r$ to retrieve points.
For each of cells spatially overlapping with $Q$ at the $l_r$ level, we retrieve $100\theta$ points from its data, which is equivalent to retrieve $\theta/\alpha$ percent of data from its sample buffer.
In each incremental update, the retrieval repeatedly retrieves a chunk of $\theta/\alpha$ percent of data from sample buffers of eligible cells.
The retrieval on a level continues until either users terminate the incremental retrieval procedure or the sample buffer is exhausted after $\frac{1}{\theta}$ rounds, in which case, we move on to the next level $[(l_r+1)\texttt{mod}\frac{1}{\alpha}]$.

Suppose $l_Q$ is the lowest pyramid level in which a single cell contains the queried spatial range.
There is a $(\alpha l_Q-\alpha)$ probability that $l_r < l_Q$ and consequently the $l_r$ level has more points irrelevant with $Q$.
In such a case, to avoid most of the irrelevant points, we will simply retrieve samples from the bottom level since it contains all of the data.

When sampling is on the bottom level, the same steps are conducted on consecutive segments of leaf cells' circular arrays.
In the case $l_r \geq l_Q$, after the retrieval has obtained points from Level $l_r$ to Level $(\frac{1}{\alpha}-1)$, the retrieval begins at the $\frac{1}{\alpha}$-th segment. Otherwise, it starts at the $l_r$-th segment.
Similarly to sample buffers, the retrieval procedure accesses the same $\theta/\alpha$ portion of points in a segment, continue, and will exhaust all eligible points after $1/\theta$ times. 
Then, the index of the next retrieved segment is $(l_r+1)\texttt{mod}\frac{1}{\alpha}$. Likewise, the retrieval continues until either users cancel or $l_r$ is reached again.

Algorithm \ref{algo:in-memory-sampling} describes the whole retrieval procedure in a time bin.

\begin{algorithm} [!htb]
\SetKwData{Left}{left}\SetKwData{This}{this}\SetKwData{Up}{up}
\SetKwFunction{Union}{Union}\SetKwFunction{FindCompress}{FindCompress}
\SetKwInOut{Input}{input}\SetKwInOut{Output}{output}
\Input{Query $Q$; $\theta$}
\Output{A random sample $S$ with $100\theta$ percent of points}
\BlankLine
Determine $l_Q$ according to the spatial query range of $Q$\; 
Initialize empty lists $S$ and $G$\;
$l_r \leftarrow$ a level randomly chosen between 1 and $\frac{1}{\alpha}$\; 
$l \leftarrow l_r$; $u \leftarrow 1$\;
\While{($u <= \frac{1}{\theta}$ or users didn't terminate)}{
     \If{$G$ is empty}{
        \eIf{$l_Q \leq l_r$ and $l < \frac{1}{\alpha}$}{
            $G \leftarrow $ sample buffers of cells that are in the $l$-th level and spatially overlapping with $Q$\;
        }{
            $G \leftarrow $ the $l$-th segments of cells that are in the leaf level and spatially overlapping with $Q$\;
        }
    }
    $u_0 \leftarrow (u $ mod $ \frac{\alpha}{\theta})==0$? $\frac{\alpha}{\theta}:u$ mod $ \frac{\alpha}{\theta}$\;
    \For{each element $b$ in $G$}{
    $s \; \leftarrow $ points satisfying $Q$ and in the $[100(u_0-1){\theta}/{\alpha}\%, 100u_0{\theta}/{\alpha}\%]$ portion of $b$\;
    $S \leftarrow S \cup s$\;
    Send $S$ to a VA system for visualization;
    }
    $u \leftarrow u+1$\;
     \If{$u$ \text{mod} $\frac{\alpha}{\theta} == 0$}{
        $G \leftarrow$ an empty list\;
        $l \; \leftarrow$ 1 + ($l$ mod $\frac{1}{\alpha})$ \;
        \If{$l == l_0$}{
            break\;
        }
    }
}
\caption{Retrieving samples in Temporal Bin $t$}
\label{algo:in-memory-sampling}
\end{algorithm}\DecMargin{1em}

\subsection{Index Update}
\label{sec:stus-append-data}

Once new data arrive, \texttt{STULL} updates its data index through finding temporal bins associated with the new points, adding the points into leaf cells and following Algorithm~\ref{algo:build-sample-buffer} to  refresh sample buffers in each associated bin.
This updating procedure applies to both existing data and new streams of data.
In general, existing data (e.g., historical logs) are well collected and curated before the visual analytics process.
Thus, its index update has sufficient time to conduct before queries, unlike streaming data, which must be timed carefully. 
Incoming data streams and their queries span more recent time ranges (e.g., a monitoring system~\cite{Magdy:2014ig} querying sensor data collected in the last ten minutes); the update procedure likely adjusts only the latest few temporal bins, which is therefore fast.

% \begin{enumerate}
% \item Find a set $M$ consisting of temporal bins whose intervals overlap with that of the new data. These bins are from the existing data index, or newly created if the temporal value of those new data exceed the existing range.
% \item Empty sample buffers of all pyramids in $M$.
% \item Insert the new data into pyramid leaf cells and follow Algorithm~\ref{algo:build-sample-buffer} to rebuild sample buffers.
% \end{enumerate}
\begin{table*}[!htb]
\caption{Evaluated datasets.}
\label{table:testing_data}
\centering
\renewcommand{\arraystretch}{1.1}
\begin{tabular}{l|l|r|r|l|r|l}
\hline
\multicolumn{1}{c|}{\textbf{Data}} & \multicolumn{1}{c|}{\textbf{Description}} & \multicolumn{1}{c|}{\textbf{\begin{tabular}[c]{@{}c@{}}Counts \\ (million)\end{tabular}}} & \multicolumn{1}{c|}{\textbf{\begin{tabular}[c]{@{}c@{}}Memory size \\ (MB)\end{tabular}}} & \multicolumn{1}{c|}{\textbf{Spatial~range}} & \multicolumn{1}{c|}{\textbf{\begin{tabular}[c]{@{}c@{}}Temporal \\ bin~counts\end{tabular}}} & \multicolumn{1}{c}{\textbf{\begin{tabular}[c]{@{}c@{}}Temporal \\ bin~interval\end{tabular}}} \\ \hline \hline
\begin{tabular}[c]{@{}l@{}}GEO\\\cite{geo1, geo2, geo3}\end{tabular} & \begin{tabular}[c]{@{}l@{}}human movement data from April, 2011\\ to August, 2013\end{tabular} & 5.8 & 3289 & Beijing, CHINA & 3 & year \\ \hline
OSP & \begin{tabular}[c]{@{}l@{}}Ohio traffic incident data from January 1, \\ 2012 to December 31, 2013\end{tabular} & 3.2 & 2279 & Ohio, USA & 4 & 6 months \\ \hline
\begin{tabular}[c]{@{}l@{}}Tweet-\\ Chicago\end{tabular} & \begin{tabular}[c]{@{}l@{}}tweets in Chicago from April 1, 2013 to \\ September 30, 2013\end{tabular} & 9.4 & 4452 & Chicago, IL, USA & 6 & month \\ \hline
\begin{tabular}[c]{@{}l@{}}Tweet-\\ US\end{tabular} & \begin{tabular}[c]{@{}l@{}}tweets across the entire US from January \\ 1, 2018 to March 11, 2018\end{tabular} & 12.4 & 4879 & USA & 11 & week \\ \hline
\end{tabular}
\renewcommand{\arraystretch}{1}
\end{table*}

\section{Unbiased Sampling Guarantee and Computational performance}
\label{sec:stus-unbiased-argument}

Unbiased sampling in \texttt{STULL} is guaranteed as a result of the index and the aforementioned sample retrieval procedure. We provide the theoretical proof of its unbiased claim as follows.
\texttt{STULL} also guarantees interactive rates, making it suitable for online sampling and incremental visualization. A formal computational complexity analysis is detailed in Appendix A.

We prove that \texttt{STULL} conducts unbiased sampling in two steps. First, we prove that 
\texttt{STULL} retrieves samples from one temporal bin without bias, and then prove for cases that use multiple bins.

For one temporal bin (e.g., a $t$-th bin), the sample retrieval procedure follows Algorithm 2 to access its pyramid and obtains $100\theta$ percent of points that satisfy $Q$ per visual update.
Recalling Algorithm~\ref{algo:in-memory-sampling}, the sample retrieval starts from a random level $l_r$, if $l_r >= l_Q$, and the bottom level otherwise.
The procedure in the case of $l_r >= l_Q$ is equivalent to the other procedure. Thus, we reduce the proof of unbiased sampling for just the bottom level.
Suppose $C_{Q}$ is a set of leaf cells that spatially overlap with $Q$.
In each leaf of $C_{Q}$, the retrieval process on average accesses $l$ segments of its circular queue, where $l={\theta/\alpha}$.
% and $n_{t,Q}$ is the total number of points satisfying $Q$ in the $t$-th bin.
For each leaf of $C_{Q}$, the equivalent procedure first accesses the $l_r$-th circular queue segment and then continues fetching data from $(l_r +1)$-th section in a clockwise order until $l$ segments are accessed.
Equation~\ref{eq:sampling-unbias-division} shows that each segment in the circular queue of a leaf has an equal chance of being selected.
Equation~\ref{eq:sampling-prob-in-one-leaf} shows that each point satisfying $Q$ in a leaf is also the same for other points satisfying $Q$. 
Therefore, points in each leaf of $C_{Q}$ are equally likely to be selected.

\begin{equation}
\begin{split}
 &P(\text{Segment $l_i$ is chosen}) = P(\text{$l_r=l_i$}) +  P(l_r \neq l_i) \times P(l_r\\
 & \text{~~~~ belongs to } l-1
 \text{ segments counterclockwise from } l_i)\\
 &= \frac{1}{1/\alpha} + (1- \frac{1}{1/\alpha})\frac{l-1}{1/\alpha -1}
 = l\alpha = \frac{\theta}{\alpha} \alpha = \theta
\end{split}
\label{eq:sampling-unbias-division}
\end{equation}

\begin{equation}
\begin{split}
&P({\text{Point } r \text{ is chosen}}) \\
&= \sum\nolimits_{l_i=1}^{1/\alpha} P(r \text{ is in the $l_i$-th segment})\times P(l_i \text{ is chosen})\\
&= \sum\nolimits_{l_j=1}^{1/\alpha} \frac{1}{\alpha} \times \theta
= \theta
\end{split}
\label{eq:sampling-prob-in-one-leaf}
\end{equation}

Second, we prove that sample points retrieved from multiple temporal bins are unbiased as well.
Suppose $Q$ requires $100\theta$ percent of points from each temporal bin in a set, $T_Q$.
Derived from Equation~\ref{eq:sampling-prob-in-one-leaf}, a point satisfying $Q$ in the $t$-th bin is selected with probability $\theta$.
Therefore, \texttt{STULL} ensures that each point satisfying $Q$ has the same selection probability $\theta$.

In conclusion, \texttt{STULL} is unbiased in selecting points satisfying a multidimensional query specification.

%\input{analyticsViaSTULL}
% \begin{table*}[!htb]
% \caption{Evaluated datasets.}
% \label{table:testing_data}
% \centering	
% \renewcommand{\arraystretch}{2}
% \begin{tabular}{|m{0.13\columnwidth}|m{0.60\columnwidth}|m{0.16\columnwidth}|m{0.16\columnwidth}|c|m{0.15\columnwidth}|m{0.16\columnwidth}|}
% \hline
% Data  & Description  & Counts (million) & Memory size (MB) & Spatial~range & Temporal bin counts&Temporal bin interval\\ \hline
% GEO \cite{geo1, geo2, geo3} & human movement data from April, 2011 to August, 2013 & 5.8&3289&Beijing, CHINA& 3& year \\\hline
% OSP   & Ohio traffic incident data from January 1, 2012 to December 31, 2013 & 3.2& 2279&Ohio, USA&4& 6 months \\\hline
% Tweet-Chicago & tweets in Chicago from April 1, 2013 to September 30, 2013& 9.4&4452&Chicago, IL, USA&6& month \\\hline
% Tweet-US & tweets across the entire US from January 1, 2018 to March 11, 2018 & 12.4&4879&USA&11& week \\\hline
% \end{tabular}
% \renewcommand{\arraystretch}{1}
% \end{table*}

\section{Evaluation}\label{sec:evaluation}
In this section, we present our experiments and results to demonstrate the effectiveness of \texttt{STULL}.

\textbf{Implementation:}
\texttt{STULL} and its two baseline approaches are built with the Microsoft .Net Framework and Visual C++~\cite{valetplatform:2018}. 
Section~\ref{sec:related_work} and Section~\ref{sec:intuition} elaborate on our baseline choice.
\begin{itemize}
    
\item \texttt{STORM}~\cite{STORM_full} is a spatial online sampling approach, using a sample-buffer equipped R-tree~\cite{r-tree:1984} to index data.
It uses the fixed-sized sample buffer design, making non-leaf cells have the same number of points in their buffers.
In our experiments, \texttt{STORM} used the \textsf{Boost} library API~\cite{Boost:2018} to build a quadratic R-tree~\cite{r-tree:1984}, and each of its sample buffers have 1024 points.

\item \texttt{RandomPath} is a variant of a spatial sampling approach~\cite{Olken1995} that traverses a tree-based spatial index to sample data.
In our implementation, points are grouped into temporal bins, and each bin uses a Quad-tree~\cite{Quad-tree:1974} to index its points spatially. 
Unlike \texttt{STORM} and \texttt{STULL}, there are no sample buffers, and all of the points are stored merely in leaf cells.
In each bin, it follows the tree-traversal based manner~\cite{Olken1995} to retrieve samples.
\texttt{RandomPath} produces an unbiased sampling, but is slower in sample retrieval.
Thus, \texttt{RandomPath} is an approach to offline sampling instead of online sampling.
% Thus, \texttt{RandomPath} is an alternative approach to sampling in place of online sampling.
\end{itemize}

\textbf{Data sets.}
Table~\ref{table:testing_data} lists the test datasets. 
Spatial distributions among the datasets are diverse. Hotspots scattered in the OSP case and concentrate at few locations the in other three.

\textbf{Environment.} We conduct all experiments on a machine with an Intel(R) Core(TM) i7-4770K CPU at 3.5GHz, 8GB main memory, and a 256GB solid state drive.

\subsection{Numerical Accuracy of Approximate Answers}

We quantified one of the advantages of unbiased sampling through the accuracy of approximate answers expressed in numbers.
The accuracy was measured by Root Mean Square Error (RMSE)~\cite{Hyndman06anotherlook}, which calculates differences between exact answers and approximate answers.

\begin{figure}[!htb]
\centering
\includegraphics[width = 0.8\columnwidth]{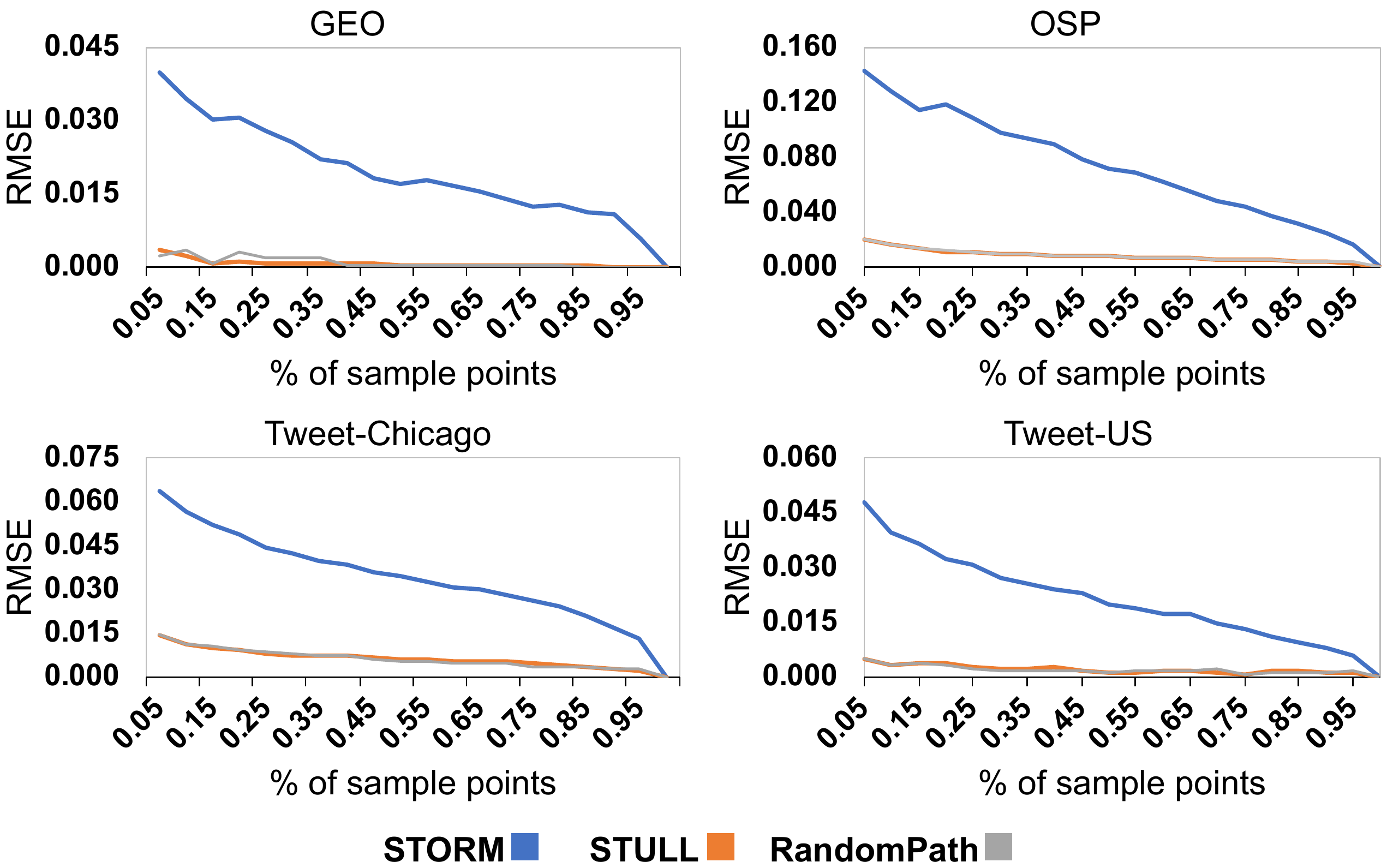}
\caption{RMSE measurement of approximate Kernel Density Estimation results.
The query requested the entire dataset.
Results were averaged over five runs.
}
\label{fig:rmse-in-mem}
\end{figure}

Regarding accuracy in the spatial dimension, we queried the Kernel Density Estimation (KDE)~\cite{doi:10.1137/1114019} results of the whole data in the geospace.
RMSE was measured on spatial bins with a KDE value no less than 0.05 on a normalized scale of 0 to 1.
Figure~\ref{fig:rmse-in-mem} shows the accuracy of approximate KDE results, compared to the exact KDE results.
Overall, RMSE values and sample sizes are inversely correlated.
At the same sample size, \texttt{STORM} has the most significant RMSE values, and the other two are almost the same or smaller.
When the sample size is 5\%, \texttt{STORM}'s RMSE value is at least twice as much as the others; and the difference decreases along with the increase of sample sizes.
Moreover, RMSE values in the OSP case are the largest at the same sample size, and nearly three times that of the other three at a particular 5\% size.

\begin{figure}[!htb]
\centering
\includegraphics[width = 0.8\columnwidth]{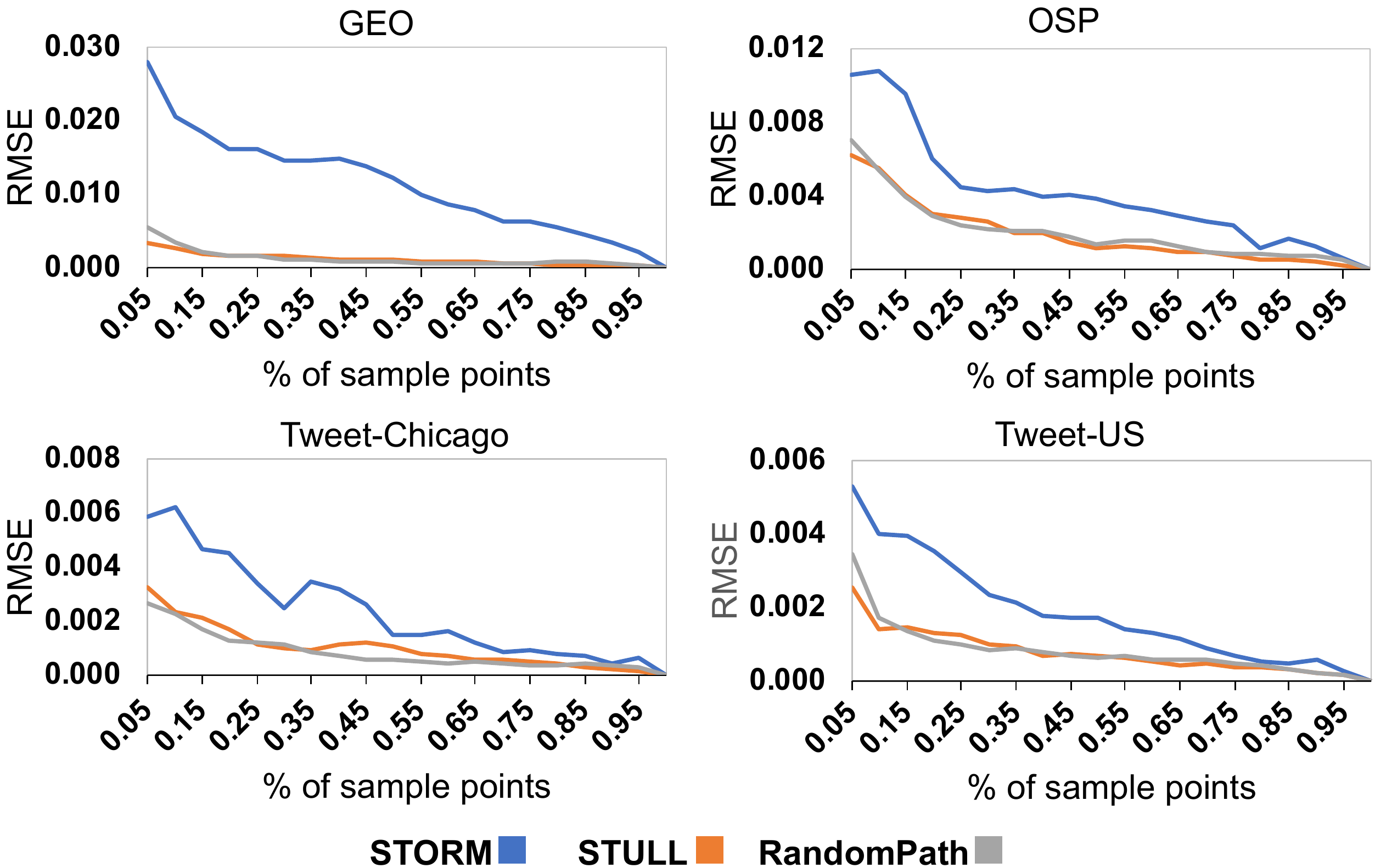}
\caption{RMSE measurement of approximate hourly distribution results.
The query pertained to every point in the entire dataset.
Results were averaged over five runs.}
\label{fig:rmse-hour-distribution}
\end{figure}

For accuracy in the temporal dimension, we queried the hourly distribution of the entire dataset.
The density value in each hour was normalized to the scale of 0 to 1.
Figure~\ref{fig:rmse-hour-distribution} shows RMSE-quantified accuracy, compared to the exact distribution.
We see that at the same sample size, \texttt{STORM} has the most significant RMSE errors.
At 5\%, \texttt{STULL}'s value is averagely 50\% less than that of \texttt{STORM}.
Furthermore, the RMSE errors in the OSP case are the largest.

\subsection{Visual Accuracy of Approximate Answers}

\begin{figure*}[!htb]
  \centering
\includegraphics[width=0.9\linewidth]{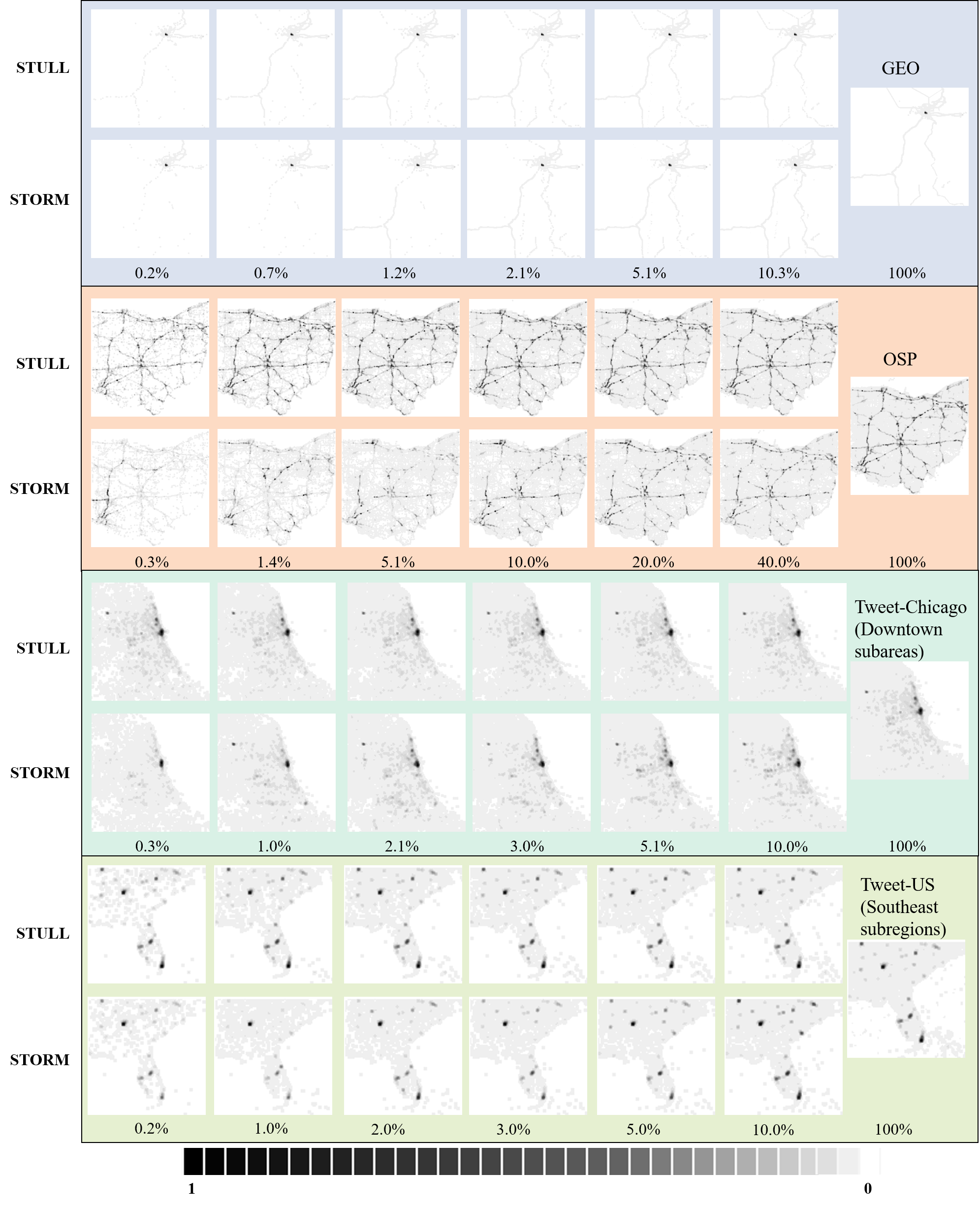}
  \vspace{-1.5mm}
  \caption{Comparison of spatial heatmaps generated by the two approaches.
%   Exact indicates heatmaps encoding exact spatial distributions, and 
A number below a heatmap indicates number of sample points selected for approximate distributions. 
At the bottom is the gray-scale colormap with 32 shades.
%   \textcolor{blue}{Heatmaps of \texttt{RandomPath} are not shown here since it is an unbiased approach.}
  }
  \label{fig:vis-compare-heatmap}
  \vspace{-3mm}
\end{figure*}

To show the impacts of unbiased sampling on the accuracy of approximate answers expressed in visualizations, we compared them in the spatial and temporal scenarios respectively. 

Figure~\ref{fig:vis-compare-heatmap} shows incremental visualization of approximate spatial heatmaps~\cite{malik2014proactive} created by \texttt{STORM} and the unbiased-guaranteed \texttt{STULL} respectively.
Overall, heatmaps of both approaches progressively get closer to the visual appearances of the exact ones.
At a smaller sample size, both heatmaps have perceptible differences in low-density areas since these areas have fewer points selected;
When sample sizes exceed certain numbers, heatmaps of the both approaches display indiscernible visual appearances.
At the same sample sizes, heatmaps generated by \texttt{STORM} are perceived as presenting more visual differences from the exact heatmaps than the other; likewise, \texttt{STULL} uses less samples to generate heatmaps that are indiscernible from the exact ones in terms of human perception.
In the incremental updates, hotspot (densities values at least 0.5) distributions in the \texttt{STULL}'s heatmaps keep constant without discernible changes, whereas hotspots in the \texttt{STORM} cases have noticeable changes when the sample sizes are smaller, e.g., the heatmap with 1.4\% points and the heatmap with 5.1\% points in the OSP case.

\begin{figure}[!htb]
    \centering
    \includegraphics[width=\linewidth]{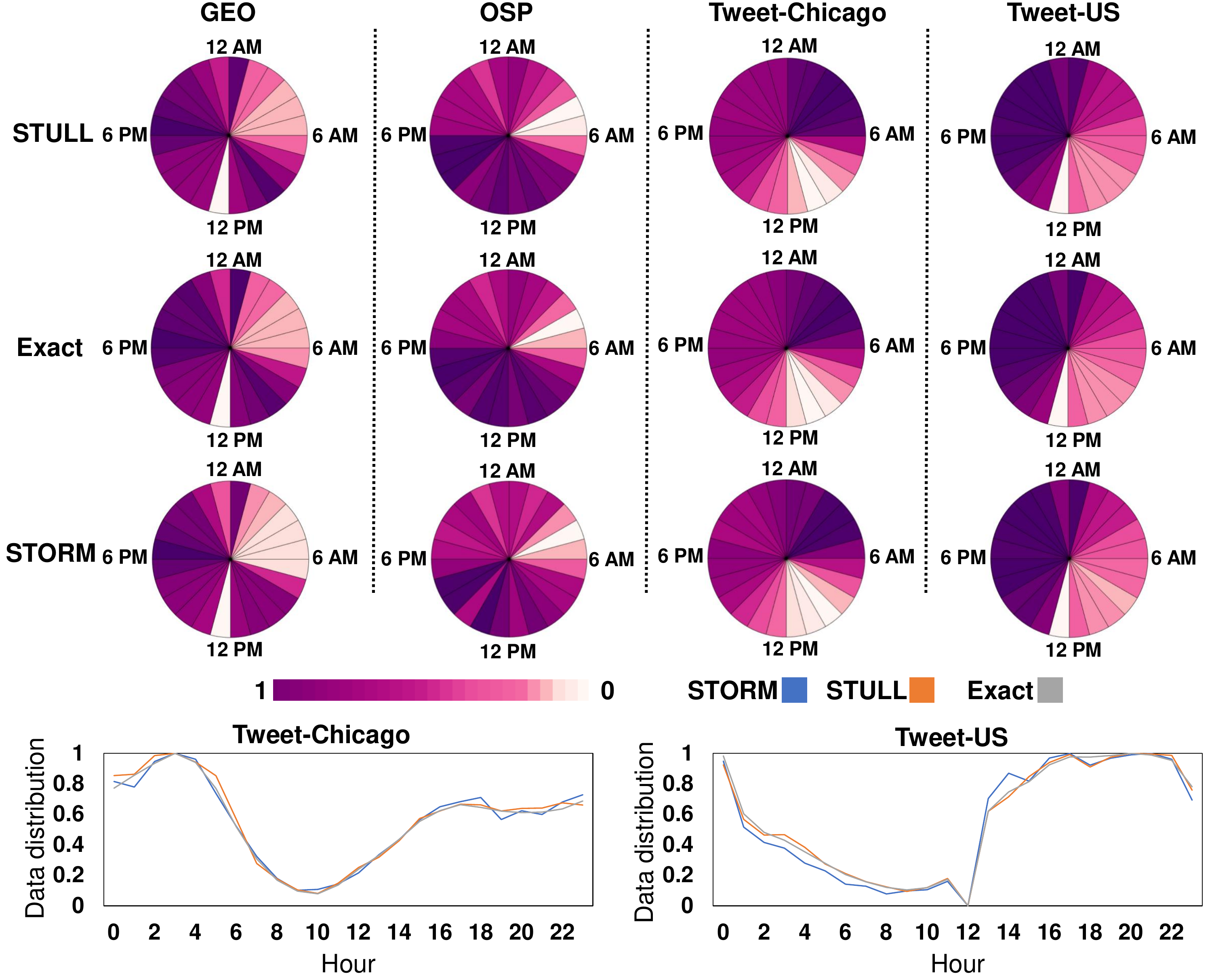}
    \caption{Pie charts showing normalized hourly distribution of the entire data. Each slice denotes a hour. 
    These approximate charts are generated with 0.1\% points of being selected. The color legend has 32 color shades. Line charts show the two datasets having closer visual appearances in the pie chart views.
    The result is one-time run. }
    \label{fig:clock_view}
\end{figure}

As for the temporal dimension, Figure~\ref{fig:clock_view} compares pie charts encoding the hourly distribution of points selected by \texttt{STULL} and \texttt{STORM}.
Overall, pie charts associated with \texttt{STULL} have less visual differences from the exact ones than those with \texttt{STORM}.
In the OSP case, point densities between 12 PM and 6 PM extrapolated from the \texttt{STORM}-supported chart clearly disagree with that of the exact one.
This is also true of the GEO case, where densities between 11 PM and 6 AM extrapolated from \texttt{STORM}'s chart have obvious discrepancies.
\texttt{STORM} also presents light but discernible color differences between 6 PM and 12 AM in the Tweet-Chicago case and between 1 PM and 5 PM in the Tweet-US case.

% \begin{figure}[!htb]
%     \centering
%     \includegraphics[width=\linewidth]{hour_dist_new.png}
%     \caption{Line charts showing normalized hourly distribution of the entire data. Each tick denotes a hour. 
%     These approximate charts are generated with 0.1\% points of being selected.
%     The result is one-time run. }
%     \label{fig:clock_view_line}
% \end{figure}

% \begin{figure}[!htb]
%     \centering
%     \includegraphics[width=0.95\linewidth]{incre-heatmap.png}
%     \vspace{-1em}
%     \caption{Incremental updates of heatmaps following the progressive retrieval of sample points by \texttt{STULL}.}
%     \label{fig:increvis-heatmap}
% \end{figure}

\subsection{Latency of Incremental Updates} 

We measured incremental sample retrieval latency in multiple scenarios.

\begin{figure}[!htb]
\centering
\includegraphics[width = 0.8\columnwidth]{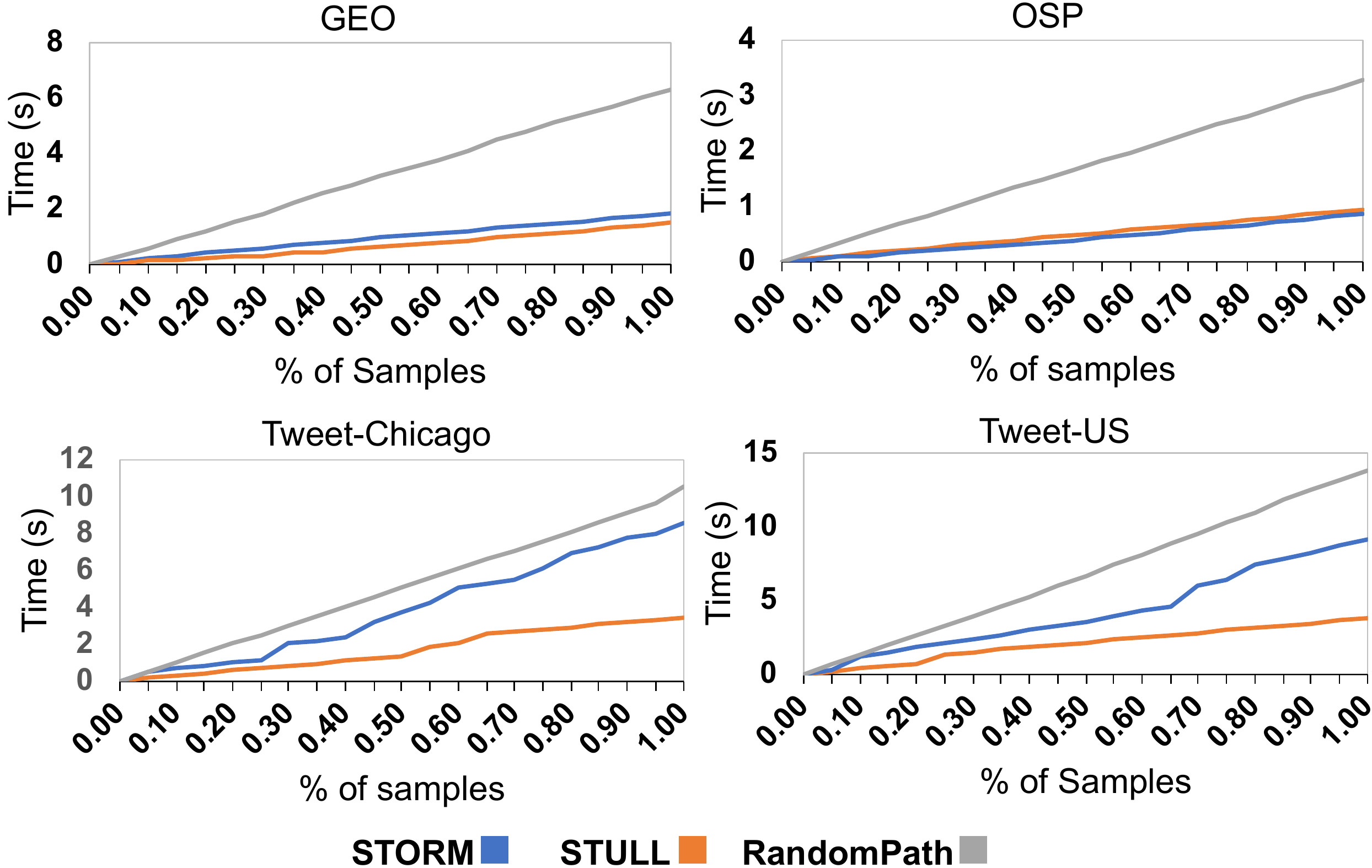}
\caption{Time measurements (in seconds) to retrieve samples in an in-memory setting. 
The query requires the entire data.
\texttt{STULL} has $\alpha=0.25$, and \texttt{RandomPath} has at most 4 levels in each of its Quad-trees.
Results are averaged over 5 runs. 
}
\label{fig:querytime-historical}
\end{figure}

\begin{figure}[!htb]
\centering
\includegraphics[width = 0.8\columnwidth]{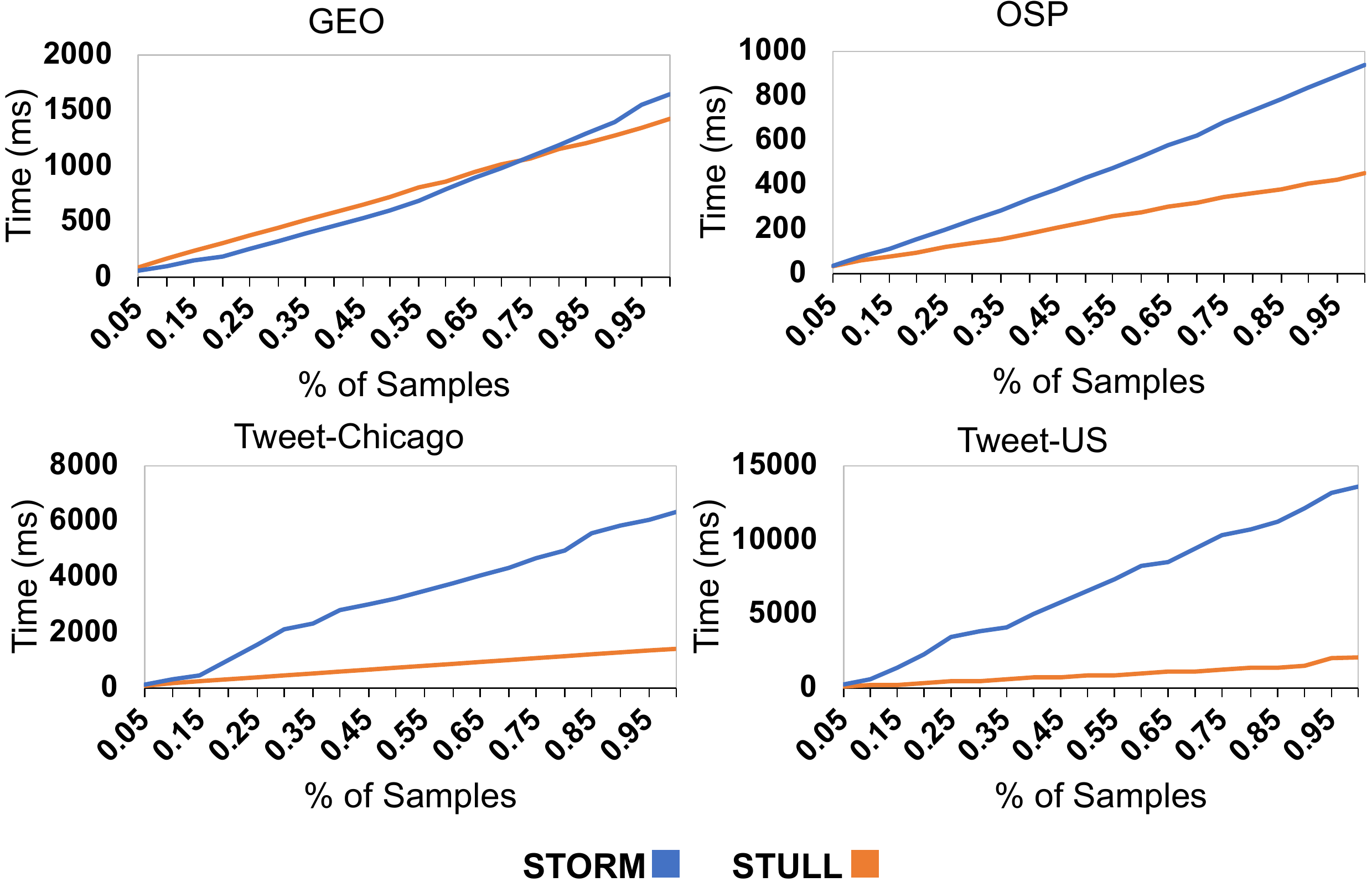}
\caption{Time measurements (in milliseconds) to retrieve samples from an in-memory data index.
Queried temporal ranges are, 2012 for OSP, 2011-2012 for GEO, 2013/04-2013/06 for Tweet-Chicago, and 2018/01/01-2018/02/04 for Tweet-US.
\texttt{STULL} has $\alpha=0.25$.
Results are averaged over five runs.
% \textcolor{blue}{Results of \texttt{RandomPath} are not shown here since it adopts the same way to determine the number of sampling points per temporal bin.}
}
\label{fig:querytime-historical-half}
\end{figure}

\begin{figure}[!htb]
    \centering
    \includegraphics[width=0.8\columnwidth]{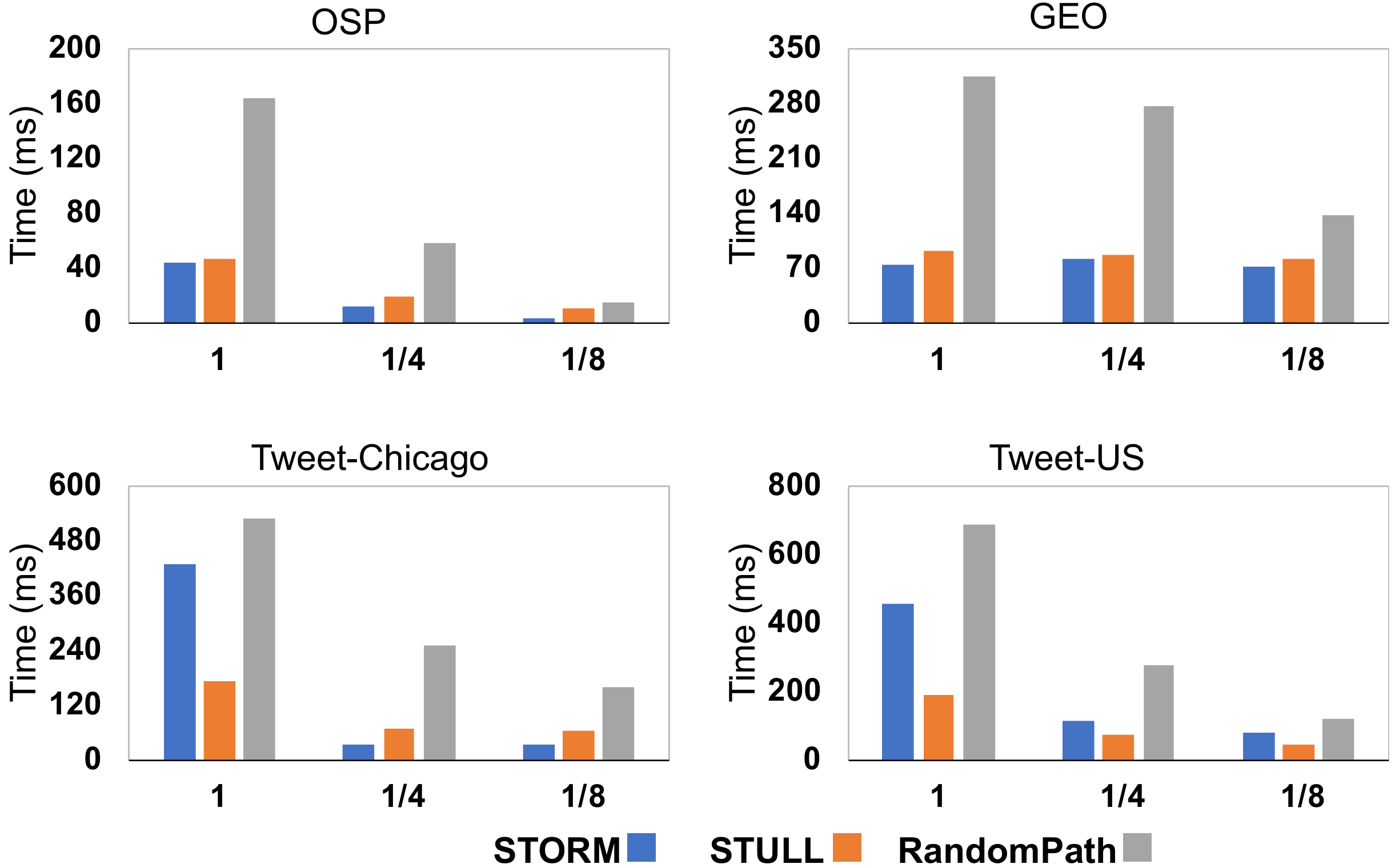}
    \caption{Average time per incremental update.
    Each incremental update retrieved 5\% points.
    Numbers below a bar indicate queried spatial range, 1 for the whole spatial extent, $1/4$ for a quarter of the whole extent, and $1/8$ for a one-eighth. For \texttt{STULL}, $\alpha=0.25$. Each of \texttt{RandomPath}'s Quad-trees has at most 4 levels.
    }
    \label{fig:retrieve_time_per_update}
\end{figure}

First, a series of experiments were conducted when data indexes were in-memory.
Figure~\ref{fig:querytime-historical} shows the time spent progressively retrieving samples for a query that queried the entire data.
It shows that \texttt{STULL} can retrieve a sample of 5\% data in less than 250ms, and the entire dataset is retrieved in 1 to 4 seconds, depending on the data volume.
Both \texttt{STULL} and \texttt{STORM} have almost the same retrieval latencies, which are overall shorter than \texttt{RandomPath}.
On average, at the same sample size, \texttt{STULL} saved at least 60\% of the time used by \texttt{RandomPath}.
Figure~\ref{fig:querytime-historical-half} presents the same time measurement for a query requiring partial temporal ranges.
It shows that \texttt{STULL} is faster because it retrieves from partial temporal bins, but \texttt{STORM} indexed points only in the spatial dimension and needs to access the entire dataset to filter out points in a temporal sub-range.
Figure~\ref{fig:retrieve_time_per_update} shows averaged sample retrieval latency in one incremental update for queries requiring various spatial ranges.
It shows that \texttt{RandomPath} takes a longer time than the other approaches.
At the same sample size, \texttt{STULL} takes less than 35\% of the time used by \texttt{RandomPath}.
Moreover, when queried spatial extents expand, \texttt{STULL} remains almost the same, whereas \texttt{RandomPath} changes significantly.

\begin{figure}[!htb]
    \centering
    \includegraphics[width=0.8\columnwidth]{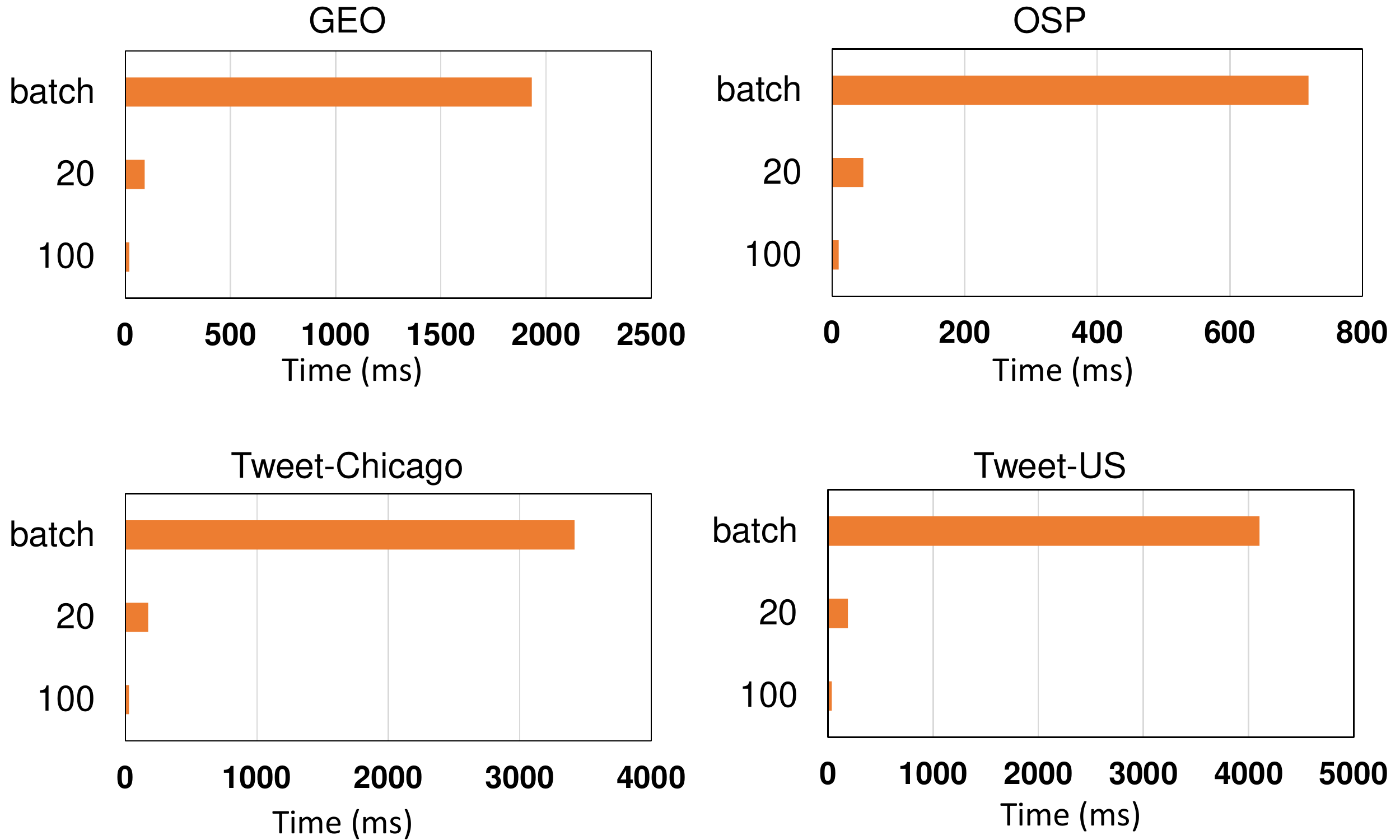}
    \caption{Averaged sample retrieval latency per incremental update. The query required the entire data.
    In the y-axis, batch indicates time to retrieve the entire dataset, 20 indicates incremental visualization has 20 updates in total and retrieves 5\% point per update; likewise, 100 indicates 1\% per update and 100 updates in total. For \texttt{STULL}, $\alpha=0.25$.
    Results are averaged over five runs.}
    \label{fig:retrieve_time_per_update_beta}
\end{figure}

\begin{figure}[!htb]
    \centering
    \includegraphics[width=0.8\columnwidth]{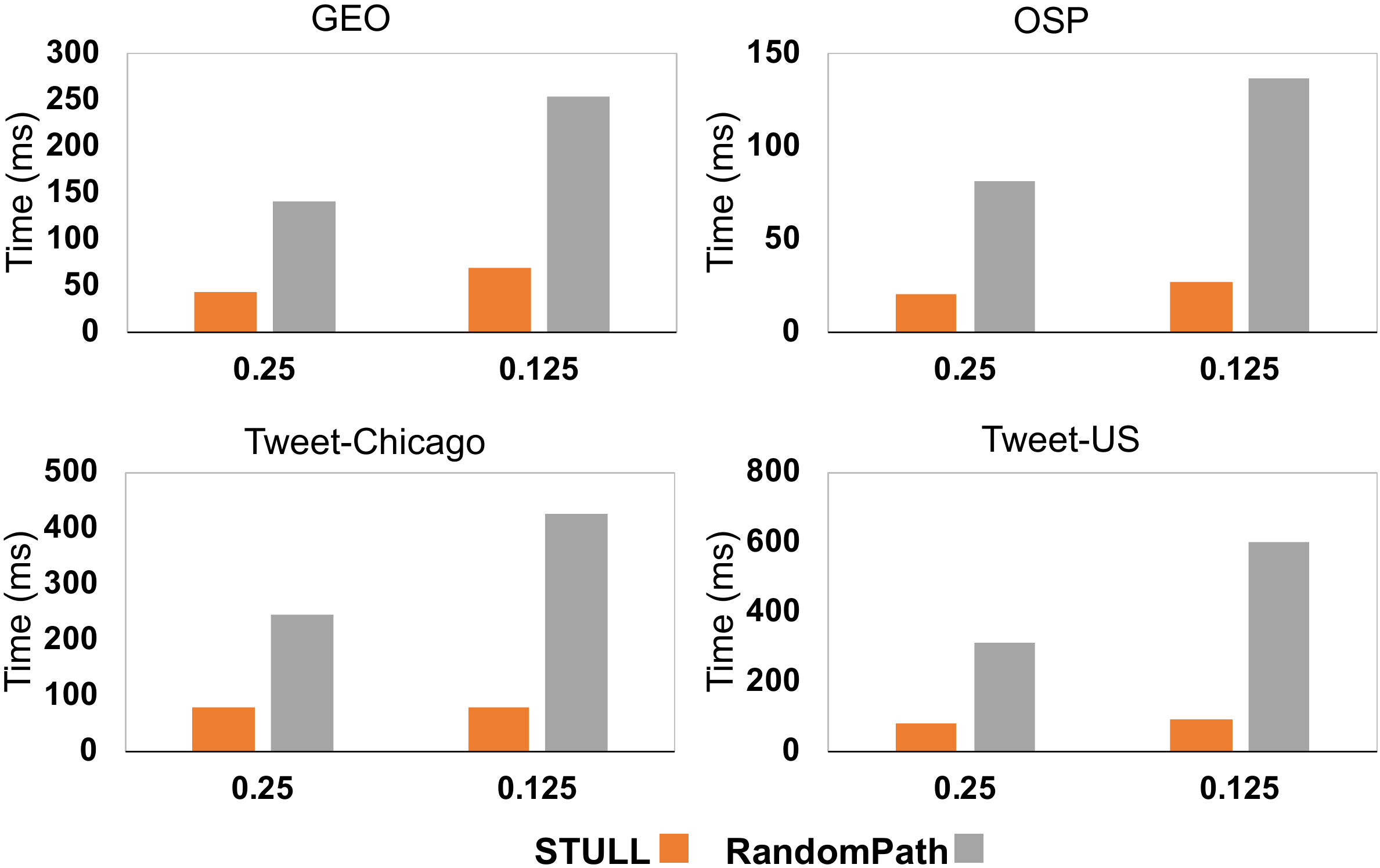}
    \caption{Averaged sample retrieval latency per incremental update. Each incremental update retrieved 2.5\% points. 0.25 indicates that $\alpha$ of \texttt{STULL} is 0.25, and \texttt{RandomPath}'s Quad-tree index has no more than 4 levels.
    Likewise, 0.125 indicates $\alpha$ = 0.125, and a Quad-tree index has at most 8 levels.
    Results are averaged over three runs.}
    \label{fig:retrieve_time_per_alpha}
\end{figure}

Second, as $\alpha$ and $\theta$ are essential parameters for \texttt{STULL}, we measured sampling latency under various values of the two.
Figure~\ref{fig:retrieve_time_per_update_beta} shows averaged sample retrieval time per update under different number of points retrieved per update.
It shows that the average time per update is almost proportional to the number of points required in each update. 
Figure~\ref{fig:retrieve_time_per_alpha} shows sample retrieval latency regarding $\alpha$.
It shows that \texttt{STULL} saves at least 68\% of the time used by \texttt{RandomPath} under the $\alpha=0.25$ settings and at least 70\% of the time under the $\alpha=0.125$ settings.
In addition, the latency of \texttt{STULL} stays the same or increases no more than 35\% if $\alpha$ reduces from 0.25 to 0.125, compared to \texttt{RandomPath}, which increases more.

\begin{figure}
    \centering
    \includegraphics[width=0.8\columnwidth]{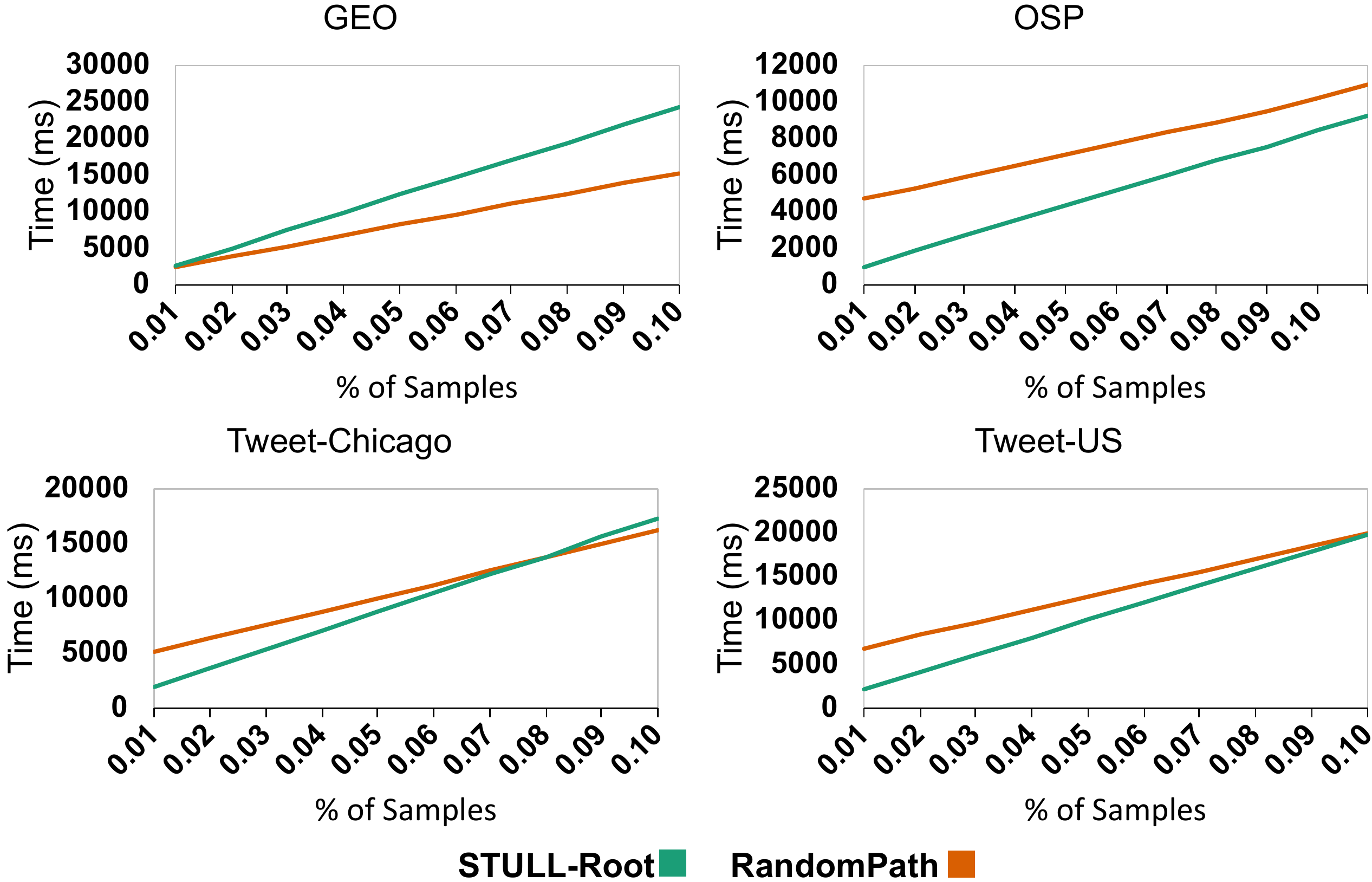}
    \caption{Latency to retrieve samples from disk-resident indexes for a query requiring the entire data. 
    Each incremental update obtains 5\% points.
    STULL-Root refers \texttt{STULL} started retrieval from the pyramid root in each temporal bin.
    For \texttt{STULL}, $\alpha=0.25$. For \texttt{RandomPath}, each Quad-tree has at most 4 levels.
    Results are averaged over three runs.
    % \textcolor{blue}{explain why STORM is not listed here.}
    }
    \label{fig:disk_cache_retrieval}
\end{figure}

Lastly, we measured sample retrieval latency when an index was stored in hard drives. Figure~\ref{fig:disk_cache_retrieval} compares retrieval time between \texttt{STULL} and \texttt{RandomPath}.
When sampling 5\% points for the initial visual update, at least 62\% of the time needed by \texttt{RandomPath}, averagely 3.2-4.7 seconds, is saved by \texttt{STULL} if it starts the retrieval at the pyramid root. 
But in the GEO case, it takes almost the same time as \texttt{RandomPath}.
GEO points are extremely concentrated in a few leaves. As a result, the time needed for \texttt{RandomPath} to load points from other leaf cells is negligible.  

\subsection{Latency on Index Creation and Update}

Computational complexity analysis (in Appendix A) shows that index creation and update are impacted by $\alpha$.
Thus, we conducted experiments on historical data and streaming data with different $\alpha$.
Table~\ref{tab:data-organization-time} shows that the average time to index historical logs is inversely correlated with $\alpha$. It indicated that latency doubled when $\alpha$ drops from 0.25 to 0.125.
For streaming data, Table~\ref{tab:data-insert-time} shows the average time to insert new arrivals of 5000 tweets, with the assumed streaming data arrival rate around 1000$\sim$4000+ per second~\cite{Mercury:2014,STORM_full}. 
This is a simulated experiment where we randomly select 5000 points from a temporal bin and measure the time required to add these data into the same bin. The recorded time is a sum of the time to add data to the pyramid and time to build sample buffers.
We can see that the insertion takes 75\% more time in the Tweet-US case and less than 33\% in other cases.

% \begin{table}[!htbp]
% \caption{Time measurements (in seconds) using \texttt{STULL} to index data. Results are averaged over five runs.}
% \label{tab:data-organization-time}
% \centering
% \renewcommand{\arraystretch}{2}
% \begin{tabular}
% {|m{0.17\columnwidth}|m{0.14\columnwidth}|m{0.1\columnwidth}|m{0.14\columnwidth}|m{0.14\columnwidth}|}
% \hline
%  & GEO & OSP & Tweet-Chicago & Tweet-US\\\hline
% %\texttt{STORM} & 63.0 & 112.9 &212.6 & 308.7\\\hline
% %\texttt{RandomPath} & 62.0 & ~86.3 & 193.7 & 179.0 \\\hline
% $\alpha=0.250$ & ~~76.142 & ~~58.461 & 181.493 & 163.234\\\hline
% $\alpha=0.125$ & 172.233 & 206.146 & 243.323 & 227.889\\\hline
% \end{tabular}
% \renewcommand{\arraystretch}{1}
% \end{table}

\begin{table}[!htbp]
\caption{Time measurements (in seconds) using \texttt{STULL} to index data. Results are averaged over five runs.}
\label{tab:data-organization-time}
\centering
\renewcommand{\arraystretch}{2}
\begin{tabular}{|r|r|r|r|r|}
\hline
\multicolumn{1}{|l|}{} & \multicolumn{1}{c}{\textbf{GEO}} & \multicolumn{1}{|c|}{\textbf{OSP}} & \multicolumn{1}{l|}{\textbf{\renewcommand{\arraystretch}{1}\begin{tabular}[c]{@{}l@{}}Tweet-\\Chicago\end{tabular}}} & \multicolumn{1}{l|}{\textbf{\renewcommand{\arraystretch}{1}\begin{tabular}[c]{@{}l@{}}Tweet-\\US\end{tabular}}} \\ \hline
$\alpha=0.250$ & 76.142 & 58.461 & 181.493 & 163.234 \\ \hline 
$\alpha=0.125$ & 172.233 & 206.146 & 243.323 & 227.889 \\ \hline
\end{tabular}
\renewcommand{\arraystretch}{1}
\end{table}

% \begin{table} [!htb]
% \caption{Time measurements (in milliseconds) for \texttt{STULL} to insert 5000 points into the existing data index. Results are averaged over five runs.}
% \label{tab:data-insert-time}
% \centering
% \renewcommand{\arraystretch}{2}
% \begin{tabular}
% {|m{0.14\columnwidth}|m{0.14\columnwidth}|m{0.14\columnwidth}|m{0.14\columnwidth}|m{0.14\columnwidth}|}
% \hline 
%   & OSP & GEO & Tweet-Chicago & Tweet-US\\\hline
%  $\alpha$ = 0.250 & 153.103 & 218.732 & 268.735 & 190.614\\\hline
%  $\alpha$ = 0.125 & 203.089 & 260.397 & 346.852 & 334.355\\\hline
% \end{tabular}
% \renewcommand{\arraystretch}{1}
% \end{table}

\begin{table} [!htb]
\caption{Time measurements (in milliseconds) for \texttt{STULL} to insert 5000 points into the existing index. Results are averaged over five runs.}
\label{tab:data-insert-time}
\centering
\renewcommand{\arraystretch}{2}
\begin{tabular}{|r|r|r|r|r|}
\hline
\multicolumn{1}{|l|}{} & \multicolumn{1}{c}{\textbf{GEO}} & \multicolumn{1}{|c|}{\textbf{OSP}} & \multicolumn{1}{l|}{\textbf{\renewcommand{\arraystretch}{1}\begin{tabular}[c]{@{}l@{}}Tweet-\\Chicago\end{tabular}}} & \multicolumn{1}{l|}{\textbf{\renewcommand{\arraystretch}{1}\begin{tabular}[c]{@{}l@{}}Tweet-\\US\end{tabular}}} \\ \hline
$\alpha=0.250$ & 218.732 & 153.103 & 268.735 & 190.614 \\ \hline 
$\alpha=0.125$ & 260.397 & 203.089 & 346.852 & 334.355 \\ \hline
\end{tabular}
\renewcommand{\arraystretch}{1}
\end{table}
\section{Discussion}
\label{sec:discussion}

We validated the \textbf{importance of the unbiased guarantee for incremental visualization} through designing experiments that measured numerical and visual accuracy between approximate and exact answers.
RMSE results (Figure~\ref{fig:rmse-in-mem} and Figure~\ref{fig:rmse-hour-distribution}) show that compared to \texttt{STORM}, unbiased sampling reduces both spatial and temporal distribution errors by at least 50\%, given a sample set of 5\% points.
The same accuracies between \texttt{STULL} and \texttt{RandomPath} confirm that our online-manner approach can ensure the same sampling quality as a regular unbiased sampling approach. 
Figure~\ref{fig:vis-compare-heatmap} and Figure~\ref{fig:clock_view} show that unlike \texttt{STORM}, approximate spatial heatmaps and approximate hourly pie charts constructed by unbiased samples have closer visual appearances to exact answers.
Consequently, users have a higher chance of inferring high-fidelity answers from approximate visuals.
Improved accuracy on the numerical measurements and visual effects are vital for incremental visualization in terms of user uncertainty~\cite{Fisher:trust:2017, Fisher:trust:users:2017}.
Users feel uncertain about choosing trustworthy approximate answers for their decision-making.
A common solution to facilitate user evaluation of the answer reliability is the use of statistical measurements derived from numerical properties of approximate answers (e.g., Confidence Interval)~\cite{Fisher:trust:users:2017, Hellerstein:1997:OA}.
\texttt{STULL} can provide samples that have better performances in such measurements, thereby helping users reduce uncertainty and obtain confidence in choosing the best answers.
In addition, improved visual accuracy confirms that unbiased-guarantee incremental visualizations can take fewer sample points to provide visual answers equivalent to the exact answers.
This is crucial to incremental visualization, because users probably use the visualizations presented in the first few visual updates to check whether the data selection conditions in a query are correct or not~\cite{Fisher:trust:2012}.
Visualizations created from biased samples can mislead users that they issued wrong queries and need to do some correction, whereas the query specifications are correct.
As a result, users' mental efforts to use incremental visualization for data exploration and decision-making significantly increase.

Specific to \textbf{geospatial accuracy}, \texttt{STULL} is affected mainly by sample size~\cite{lohr2009sampling}.
However, \texttt{STORM} has one more factor, intrinsic spatial distributions in the data. 
In spatially clustered distribution cases (e.g., GEO), compared to \texttt{STULL}, \texttt{STORM} has light or indiscernible visual differences in hotspot areas, but is incompetent in lower-density areas.
On the other hand, in a scattered distribution case (e.g., OSP), the absence of an unbiased guarantee causes \texttt{STORM} to extend defective visual appearances to hotspots, e.g., incorrect hotspot locations.
The RMSE value (Figure~\ref{fig:rmse-in-mem}) in the OSP case is almost ten times that of the concentrated cases. 
So is the visual effect in which \texttt{STORM} defectively represented in wider spatial extents in the OSP case but misrepresented merely sparse ones in the concentrated scenes. 
We believe \texttt{STORM}'s reduced accuracy loss in spatially concentrated cases is caused by the fixed-size sample buffer. 
First, \texttt{STORM}'s spatial index,  \texttt{R-Tree} tends to create more cells in high-density areas, and fewer cells in low-density areas.
Accordingly, in a spatially concentrated case, a majority of cells are associated with scarce hotspot regions.
Since in the fixed-sized buffer design, the number of points sampled in a region is proportional to the number of its cells, 
\texttt{STORM} can retrieve more points in order to characterize hotspot areas but
pay less attention to lower-density areas.

\texttt{STULL} achieves the \textbf{temporal unbiased property} through determining the number of samples retrieved from each temporal bin proportional to the total data volume in the bin, which is a  widely used golden rule~\cite{trajstore:2010:icde,Li:2015}. Thus, we do not elaborate on it.

Our range of experiments also validate the efficiency of STULL's
sample retrieval. 
Experiments (Figure~\ref{fig:querytime-historical}, Figure~\ref{fig:querytime-historical-half}, Figure~\ref{fig:retrieve_time_per_update}) indicate that compared to \texttt{RandomPath}, \texttt{STULL} can reduce latency by at least 60\% to sample 5\% of in-memory data per incremental update despite query specification at various geospatial and temporal scales. 
As for the case in where data indexes are on disk drives, \texttt{STULL} and \texttt{STORM} load points from buffers of root cells first and continue retrieval from buffers of its descendants.
This significantly reduces the number of \texttt{STULL}'s disk I/Os needed for the first visual update, whereas \texttt{RandomPath} needs to retrieve points from most of its leaf cells, consequently forcing almost every cell to be loaded into memory, which results in an extremely slow response. 
% Since \texttt{STULL} and \texttt{STORM} merely access buffers in the top levels of pyramids, they clearly mitigates disk I/O costs and rapidly provide samples.
Thus, \texttt{RandomPath} does not satisfy the critical latency.

Our experiments show that users are able to keep \textbf{computational latency per incremental update} well under control.
Figure~\ref{fig:retrieve_time_per_update_beta} shows that \texttt{STULL} successfully controls retrieval time proportional to the number of points per visual update.
% As a result, users can reduce $\beta$ if incremental updates are expected to be quicker.
In addition, our experiments sequentially accessed temporal bins to obtain samples, which resulted in higher latency compared to an in-parallel scheme.
We leave STULL's adoption of parallel techniques to future work.

\texttt{STULL} \textbf{reduces the data index creation and update workload}, compared to \texttt{STORM}.
\texttt{STULL} indexes data spatially using pyramids for efficiency~\cite{Mercury:2014}. 
We conducted experiments to measure index creation time and confirmed the superiority of our pyramid-based approach, which is approximately 10\% faster than the R-tree based \texttt{STORM}.
% In addition, since \texttt{STORM} starts sample retrieval at the root of its index, a systematic bias issue is inevitable, since sample buffers in the top-level are retrieved prior to their descents'.
% To avoid such bias, \texttt{STORM} needs to rebuild its sample buffer routinely, which inevitably increases its maintenance workload.
Regarding streaming data, Table~\ref{tab:data-insert-time} shows that it takes less than 450ms to index 5000 points. 
Thus, inserting the new data into the existing data index and retrieving samples from the updated data index can be completed in approximately less than 500 ms for the OSP, Tweet-Chicago and Tweet-US datasets and approximately 600ms for the GEO.

\texttt{STULL} is designed for aggregation-based \textbf{spatiotemporal analytics} that assist end-users in summarizing trends and patterns of data, e.g., estimating data count per hour or evaluating averaged statistics of income per neighborhood.
Here, we demonstrate aggregation computation with samples retrieved by \texttt{STULL} and use confidence intervals~\cite{lohr2009sampling, Hellerstein:1997:OA} to estimate the proximity of generated approximate answers to exact answers.
Suppose a query calculates the average length of tweets posted in the morning. 
A sample $S$ of $n$ tweets is retrieved by \texttt{STULL}. The average length of tweets is $\Bar{v}=\frac{1}{n}\sum_{s_i\in S} v_i$, $v_i$ is the length of the $i$-th tweet. Let $c$ denotes the standard deviation of the sample estimate. Thus, the interval $[v-2c, v+2c]$ contains the exact answer with 95\% of the time.

Despite efficient sampling, \texttt{STULL} suffers from \textbf{inefficient usage of storage space}.
\texttt{STULL}'s pyramids contain all points in the bottom levels, 
in addition to $(1-\alpha)$ portion of data in non-bottom levels, whereas \texttt{Quad-Tree} does not have such extra costs.
Thus, \texttt{STULL} maximizes retrieval efficiency at the expense of data storage space. 
If data stored at non-bottom levels are not duplicated at the bottom level, \texttt{STULL} will move from the bottom level to the higher levels, and retrieve samples from these higher levels. 
Since higher levels consist of cells whose spatial ranges are quadratically larger than the queried range, we could anticipate a 3-fold increase in retrieval time. 
Although the duplicated data will take additional storage resources, our design choice has at least two benefits. 
First, \texttt{STULL} restricts the retrieval to a minimum set of spatially relevant data, per $Q$.
Secondly, when the data index is on disk,  retrieving spatially irrelevant data in the alternative option will cause more disk I/Os.

Another limitation of \texttt{STULL} is that it does not yet fully investigate a \textbf{disk-based index}.
As data volume increases, in-memory data storage becomes scarce. 
A hybrid index of both in-memory and disk-resident data is essential to overcome the memory shortage~\cite{Benzaken:2011dn}.
For convenient disk-based data storage and retrieval, STORM~\cite{STORM_full} uses the fixed-sized design and sets the space usage of a sample buffer as equivalent to the size of a disk block, resulting in each cell has the same number of data cached in its sample buffer.
Thus, if one sample buffer is needed, \texttt{STORM} loads the corresponding disk block into memory, retrieves all data stored in that block, and removes the block from memory after use.
But in \texttt{STULL}, the proportionally-sized design causes sample buffers to have various sizes.
Consequently, it is common for one sample buffer to involve multiple blocks, with one of these blocks only partially filled. 
These partially filled blocks cause the low disk storage utilization and slow down disk I/O as well.
We leave this to future work. 
\section{Conclusion and Future Work}
\label{sec:conclusion}

This paper presents an online sampling approach, \texttt{STULL}, which samples large spatiotemporal data in an unbiased manner. 
Extensive evaluations verify that \texttt{STULL} is unbiased and computationally superior over comparable online sampling approaches.
\texttt{STULL} is suitable for a range of online data exploration including visual analytics and incremental visualization.
Approximate visualizations leveraged by \texttt{STULL} increase their numerical accuracies and reduce their visual differences as compared to the exact visualizations, when compared to approaches without unbiased guarantee.

In the future, we will extend this work 
by designing a novel scheme to store our data index on hard drives.
The current implementation has comparable performance for retrieving a small ratio of points, about 10\% in our experiment, from a disk-resident data index, but is slower if more points are needed.
A novel scheme is expected to resolve this issue.

%% if specified like this the section will be committed in review mode
\acknowledgments{
This work was supported in part by
the National Science Foundation Grant CA-FW-HTF 1937036. 
Walid G. Aref acknowledges the support of the National Science Foundation under Grant Numbers  III-1815796 and IIS-1910216.
Mingjie Tang acknowledges the support of the National Natural Science Foundation of China (Grant No. 61802364).
The authors wish to thank Jieqiong Zhao, Audrey Reinert, and Luke Snyder for their editorial comments and helps. }

\bibliographystyle{abbrv-doi}

\bibliography{template}
\end{document}